\documentstyle[12pt, leqno, amsfonts]{article}
\begin{document}
\begin{titlepage}
\title
{On Quantum Mechanics in Curved  Configurational  Space}

\author{\large E.A.Tagirov}

\date{}

\begin{abstract}
Different approaches are compared to formulation of quantum mechanics
of a particle on the  curved spaces. At first,  the canonical,
quasi-classical  and  path integration formalisms  are considered
for quantization of  geodesic motion on the Rimannian configurational
spaces. A unique rule of ordering of  operators in the canonical formalism
and a unique definition of the path integral are  established and,
thus, a part of ambiguities in the quantum counterpart of geodesic motion
is removed.  A geometric interpretation is proposed for non-invariance
of the quantum mechanics on  coordinate transformations. An approach
alternative to the quantization  of  geodesic motion
is  surveyed, which starts  with the  quantum  theory  of a  neutral
scalar field.  Consequences of this alternative  approach  and  the three
formalisms of  quantization  are compared. In particular, the  field
theoretical approach generates  a deformation of the canonical
commutation relations between  coordinates  and  momenta of a prticle.
A possible cosmological consequence of the deformation is presented
in short. Key words: quantum mechanics, Riemannian space, geodesic motion,
deformation.

\end{abstract}
\end{titlepage}
\maketitle

\newcommand{\Dx}{\Delta\xi}
\newcommand{\ep}{\epsilon}
\newcommand{\qarr}{\stackrel{\cal Q}{\longrightarrow}}
\newcommand{\lc}{L^2 (\Sg_3 (t);\ \bf C;\ b^3 (t){\sqrt \omega (t)} d^3\xi)}
\newcommand{\vp}{\varphi}
\newcommand{\1}{{\bf\hat 1}}
\newcommand{\Oc}{O\left(c^{-2(L+1)}\right)}
\newcommand{\im}{{\rm i}} 
\newcommand{\Sg}{\Sigma}
\newcommand{\bgt}{\bigotimes}
\newcommand{\ptl}{\partial}
\newcommand{\Sche}{ Schr\"odinger equation\ }
\newcommand{\Schr}{Schr\"odinger representation\ }
\newcommand{\eu}{$ E_{1,3} $}
\newcommand{\euf}{E_{1,3}}
\newcommand{\rif}{V_{1,3}}
\newcommand{\ri}{$V_{1,3}$ }
\newcommand{\ov}{\overline}
\newcommand{\stc}{\stackrel}
\newcommand{\defst}{\stackrel{def}{=}}
\newcommand{\qstc}{\stackrel{\cal Q}{\longrightarrow}}
\newcommand{\h}{\hbar}
\newcommand{\beq}{\begin{equation}}
\newcommand{\nde}{\end{equation}}
\newcommand{\beqa}{\begin{eqnarray}}
\newcommand{\ndea}{\end{eqnarray}}
\newcommand{\rin}{$V_{1,n}$}
\newcommand{\rinf}{V_{1,n}}
\newcommand{\rn}{$V_n$}
\newcommand{\rnf}{V_n}
\newcommand{\al}{\alpha}
\newcommand{\be}{\beta}
\newcommand{\ga}{\gamma}
\newcommand{\om}{\omega}

{\large\bf 1. INTRODUCTION}\\
Quantum Mechanics on the Riemannian geometric background is the simplest
part of the fundamental problem of association of general relativity and
the quantum theory. In the quantum mechanics, the problem of definition of
appropriate physical observables appears in  a relatively simple form, which
emerges  quite completely in quantization of gravitation, see, for instance,
Rovelli (1999). On the other hand,
the quantum mechanics of a point--like particle  may be considered  as
a limitng case of the string dynamics. It provides also a
description of interesting  physical  models  such  as  a motion on
homogeneous spaces of some groups, see, for instance, Marinov (1995) and
C.Groshe, G.S.Pogosyan and A.N.Sissakian (1997).
An important point is that the Quantum Field  Theory in curved space-times
 which is applied  successfully to describe
fundamental processes in the early Universe is based in fact
on some quantum  mechanics of a (quasi-)particle, at least,
implicitly, see, Gibbons and Pohle  (1993) and Tagirov (1999).
At last, one may expect  that
a modification of the well--established fundamental theory, such as
quantum mechanics,  to a more general
geometrical background can reveal new features of the theory and
serve for  better understanding of it.

The problem has  a long history related  to the names of
Podolsky,  Dirac, DeWitt and other, less known theorists. Nevertheless,
it  has  not still a satisfactory unambiguous solution.  The
main approach is  based  on the idea  of quantization
of the classical Hamiltonian systems and their generalizations.
(In  the simplest expressions, the  quantization is  a map  of
a classical theory in terms of the usual functions on a phase space
to  a mathematical structure in terms of  non-commuting objects along
definite aggregates of rules ({\it formalisms}), which depends on
a small parameter $\h $ the physical value of which is the Plank constant.)

There is a number
of different formalisms of quantization and it is natural
to expect that they give  similar results for the same physical system.
Unfortunately,   it is not the case even for such an elementary system as
the point-like chargeless and spinless particle if  the configurational
space is curved. Moreover, as a rule, there are  fundamental ambiguities
even in the framework  of the same  formalism and even for a simple
class of phase spaces $ {\cal P}_{2n}  \sim R_n \bigotimes V_n $ where
$V_n$ is   {\it the $n$-dimensional Riemannian configurational space}.

In the canonical  and   path integration
formalisms, see Sections 2 and 4 respectively, the ambiguities
appear in the following two forms. The first one  is  the known
problem  of ordering  of  operators
$\hat \xi^i , \hat p_j, \quad i,j,... = 1,...,n, $ which correspond
 to the Darboux coordinates $\xi^i, p_j$ in  ${\cal P}_{2n} $
when they are substituted  into  a function $f(\xi, p)$ (say, through
a power expansion) to obtain the  corresponding  quantum observable
$\hat f$ or the path integral.  Generally, there is no leading principle
to single out  a
 certain rule of ordering  among infinitely many ones. The ambguity does not
attract much  attention in view of that all of the rules lead to the same
operator $\hat f$  up to an  additive constant
  if $f(\xi, p) = f_0 (\xi, p) + f_1 (\xi) + f_2(p) $,
where  $f_0$ is a  second order  polynomial  of the Darboux coordinates
 $\xi, p $  and   $ f_1,\ f_2 $  are  appropiate arbitrary functions.
The classical  Hamiltonians  of  the  typical problems  of the standard
quantum mechanics in the Euclidean configurational space $E_n$   are in
this class if  the preferred Cartesian
coordinates are  taken as $\xi^i$. The latter is usually  assumed with
no stipulation.  The curvilinear coordinates  are  used,
if any, {\it a posteriori}, only as a technical tool, for example, in
relation  with a symmetry of the potential. However, even in $E_n$,
as soon as  curvilinear coordinates are taken {\it as observables},
i.e. as one--half of the phase space coordinates, then $f_0$ that was
a second order polynomial  in the Cartesian coordinates  and their conjugate
momenta, fails generally to be a polynomial at all. Respectively, the
dependence of quantization on a choice  of ordering becomes actual.
In addition,  for the path integration,  there is an
ambiguity in the choice  of  the points on a lattice of integration, in
which the integrands are evaluated, see, for instance,
D'Olivo and Torres (1989) and
Section 4 below. It is a common problem  for any
geometry of the configurational space  and  system of coordinates,
again except the case of $E_n$,  Cartesian coordinates and
the quadratic  function $f_0$.

The second ambiguity consists  in that the result
of such  quantization  depends on the choice of  coordinates
$\xi^i$ in $V_n$  , that is not invariant with
respect to diffeomorphisms of $V_n$, or, {\it diffeononinvariant} even if
a rule of ordering is fixed, though the original classical theory
is  diffeoinvariant. Again, in the standard theory, the problem is
obscured  by existence of  the Cartesian coordinates. It looks as  there
were an implicit postulate  that quantization should be performed just
in these preferred coordinates. However,   what should one   do in  \rn
where the Cartesian coordinates do not exist  at all?
An attempt to answer on  the question will be given below in the Section
5 on the basis  of results of the preceding sections.

According to  Bordemann,  Neumaier and   Waldmann (1998), the
deformation quantization    in the framework of the  Fedosov
formalism, see Fedosov (1994), leads   to  diffeoinvariant
quantum mechanics in $V_n $.  However, this result is obtained
 by using  of a particular  rule of ordering, namely, Weyl's one.
Thus, at least, the ambiguity in ordering  apparently retains.

Geometric quantization in the Blattner--Costant--Souriau formalism,
see, for instance, Abraham and Marsden (1978) and \'Sniatycki (1978),
is reduced
to the  quasi-classical approach  by Pauli--DeWitt (Pauli, 1950--1951,
DeWitt, 1957)  for the  simple case  under our consideration.
The formalism is diffeoinvariant and includes no ordering procedure, but
it is  approximate {\it ab initio} because it starts with
an Anzatz where the (unknown) quantum propagator  is substituted by
the quasi-classical one.

Among other approaches to quantization  on \rn,  it is worth  to mention
the one  based on embedding \rn to an Euclidean space of a greater
dimension and using the Cartesian coordinates in it
(Ogawa,  Fuji  and Kobushkin, 1992 ). And, at last,  the present author
develops an approach to quantum  mechanics of a particle  in \rn,  which
is an  alternative to quantization of mechanics and may be called
the quantum--field--theoretical one, or {\it the QFT-approach}
 (Tagirov, 1990, 1992, 1996, 1999). It reproduces
quantum mechanics in the general \rin in a diffeoinvariant and
ordering-independent form  as
the quasi-non-relativistic asymptotic of a  quasi-one-particle sector
on an appropriate Fock space for  the quantized neutral scalar field.
(In the paper by Tagirov (1996), the  field  of spin 1/2 is considered
but the result  needs some refinement and justification along the lines
of Tagirov (1999) and Section 6 of the present paper).
Thus, in this approach,  the canonical quantization procedure is
shifted  from the particle phase space to the quantiazation  of
the field.  The diffeoinvariant analogs of
the operators $\hat \xi, \hat p$ mentioned above prove to  satisfy
a deformation\footnotemark[1]
of the canonical commutation relations such that
the position operators mutually do not commute; of course,
the conjugate momenta are also  mutually non-commutative.
The deformation parameter  is  $c^{-2}$.
Just this and other  curious results of the
approach  stimulated the present author's interest to the state of art
in the traditional approaches to quantum mechanics in \rn.

In the present paper the three historically  first formalisms  of
quantization, the canonical, quasi-classical ones  and the path integration,
will be considered in application to the  geodesic motion in configurational
space \rn with the general time-independent {\it metric tensor}
$\om_{ij} (\xi)$. The latter means that {\it the  space-time} is
$ \rinf \sim R_1 \bigotimes \rnf$.
The Hamilton operators arising in the three formalisms
are compared in a certain approximation in which they should come
to the same Hamilton operator. This condition  distinguishes
 a unique rule of ordering of the primary observables operators
for the canonical and path integration formalisms and gives
an unambiguous prescription for the latter.  (Along the
reasons mentioned above,  these two formalisms are considered below
as "more exact" ones with respect to the quasi-classical one.)

We postpone the  deformation quantization approach and
embedding of \rn for more serious special consideration though  use
the general idea on deformation of the Poisson brackets in a formulation
of postulates of canonical  operator   formalism in Section 2.

In Section 2,  it will be shown that, for the  canonical quantization
of the geodesic motion in \rn, the freedom in the choice of ordering rules
is reduced to a one-parametric set  in each fixed  system of coordinates
$\{\xi^i \}$. Since diffeomorphisms of \rn are determined by $n$ arbitrary
$C^\infty$--functions, one may say figuratively   that the overall
arbitrariness is "$1+\infty^3$-dimensional" in this case.

In Section 3,  it is shown that the one-dimensional part of  the
arbitrariness can be  removed by condition  of coincidence
of the canonical Hamilton operator with that by  DeWitt(1957)
in a certain approximation.

In Section 4,  the  path integral for the quantum propagator of geodesic
motion  is constructed so that the phase of the integrand is
proportional to the classical action  and the  Hamilton operator generating
the propagator  coincides with  DeWitt's one in the same approximation,
as in the canonical case. This fixes the same rule of ordering  of
the primary  operators as in Section 3  and unambiguously   determines
that   the integrands should be  evaluated  at the  nodes of the lattice
of integration.

In  Section 5,  the  obtained  solution of the problem of ordering
is discussed and a  possible  explanation   of
the diffeononinvariance of the canonical quantum mechanics in \rn
is given.

In Section 6,  a survey of main results of the mentioned above
QFT-approach is given and compared with the results
of quantization of mechanics.

The paper adopts the so-called  heuristic  (or, naive) level of
mathematical rigor: many definitions and relations need further refinement
to have an exact meaning. It is expected that the latter can be
achieved if physically sensible  results  appear  at  our
imperfect level.\\

{\large\bf 2. CANONICAL QUANTIZATION OF GEODESIC MOTION IN THE RIEMANNIAN
CONFIGURATIONAL SPACE}\\

{\bf 2.1. HAMILTON THEORY OF GEODESIC MOTION}\\

To emphasize a relation of the system under consideration to  general
relativity,  let us start  with geodesic lines  in the generic
$(1+n)$-dimensional Riemannian space-time \rin  of the Lorentz signature
$-n+1$.
Let \ $ x^\al, \quad (\al, \be,... = 0, 1,..., n)  $ be
some coordinates in  \rin,   and
$ t, \xi^i, (\quad i, j,... = 1, 2, ... , n)$ be
normal  Gaussian  coordinates generated  by  the normal geodesic translation
of  a given Cauchy  hypersurface  $\Sg \equiv \Sg (t_0)$ and
some coordinates $\xi^i$ on it. The metric form is
\beqa
ds^2 &=& g_{\al\be} dx^\al dx^\be \label{g} \nonumber \\
&=& c^2 dt^2 - \om_{ij} (t, \xi) d\xi^i d\xi^j \quad t\in [t_0,\ t_1).
\label{om}
\ndea
(The range where  the coordinates $ t, \xi^i $ and  the representation
of the metric  (\ref{om}) are  valid  is indicated,  for instance,
  by Destri, Maraner and Onofri (1994), Section 2.)

The space--time geodesic lines  are  extremals  $x^\al = x^\al (s)$
of  the action functional
\beq
W = - mc \int_{s_1}^{s_2}\, ds \sqrt{g_{\alpha\beta}
\frac{dx^\alpha}{ds} \frac{dx^\beta}{ds} }\ \defst \
\int_{s_1}^{s_2}\,L^\prime\,ds = \int_{t_1}^{t_2}\,L\,dt ,
\label{w}
\nde
which satisfy  the  following constraint:
\beq
g^{\alpha\beta} (x) p_\alpha p_\beta = m^2 c^2,  \label {gpp}
\nde
where  $p_\alpha $ are the generalized  momenta
\beq
p_\gamma (s) \defst \frac{d L^\prime}{d (dx^\ga/ds)} =
- mc \, \left(g_{\alpha\beta}
\frac{dx^\alpha}{ds} \frac{dx^\beta}{ds}\right)^{-1/2}\, g_{\gamma\delta}\,
\frac{dx^\delta}{ds}.
\nde
The  canonical quantization as a map   $ Q $  of  functions on a
phase space  $ {\cal P} = {\cal P}_{2n+2} \sim T^*\rinf $   to
operators acting on a Hilbert space  $\cal H$,
(see a more exact definition below)
can be applied to this diffeoinvariant  system  with
constraint (\ref{gpp}).  However,  it would  be a map  on operators acting
 on the space\footnotemark[2]
${\cal H} \sim L^2 (\rif; {\Bbb C}; \sqrt g d_4 x)$ which
cannot be interpreted  as a space of states  of  a real
particle specified  by a position  in the configurational
space,  see  Tagirov (1999).
For the standard   probability  interpretation in the
Schr\"odinger representation,   the operators of observables  should
be defined on $L^2 (\Sg; {\Bbb C}; \sqrt \om d^n x)$.
It is realized   by quantization of the reduced Hamiltonian system
in which one solves  the constraint  (\ref{gpp})  at the classical
level. To this end,   one represents (\ref{gpp})  in the form
\beq
    \left(p_0 + mc \sqrt{1 + \frac{2H_0}{mc^2}}\right)
\left(p_0 - mc \sqrt{1 + \frac{2H_0}{mc^2}}\right) = 0,
\quad (x^0 \equiv ct),
\label{gpp2}
\nde
where
\beq
H_0 \equiv H_0 (\xi, p; t) \defst
\frac{1}{2m}\om^{ij}(\xi; t) p_i p_j  \label{ham0}
\nde
 For the nonradiating and spinless particle ( just  only it
moves along a geodesic line  in the Minkowsky space-time \rin), we shall take,  as
usually, the solution of \ref{gpp2} with respect to  $p_0$  such that
$p_0 > 0$. Then, in the  theory with the constraint thus  resolved,
the Hamilton function  will be
\beq
H(\xi, p) = mc^2 \sqrt{1 + \frac{2H_0}{mc^2}}. \label{ham}
\nde
There is an interesting intermediate approach
 of Gitman and Tyutin (1990) in which   both  the solutions of constraint
(\ref{gpp2}) are used (in \eu) through introduction of
a special observable "the sign  of $p_0$". This leads to a state space
consisting of two $L^2 (\euf; {\Bbb C};  d_3 x ) , \quad  \{x^i\} \in \euf $,
which describe particles and antiparticles respectively (being neutral,
they are  identical).  Gavrilov and Gitman  (2001) have  extended
the approach
to the  case of \ri. However,  a remark arises  concerning this work,
which  will be made at the end of the present Section, near the formula
(\ref{kin}).

The non-reduced and reduced formalisms differ in that,
in the former case, a time-like  coordinate
$x^0$  is included to the set of observables whereas, in the latter
case,  the variable $t$ is  an evolution parameter.
In the classical theory,  these  formalisms are  physically equivalent
versions of the same theory; however, quantization of them  leads
to different theories.

In the reduced formalism,  observables for  fixed $t$ are functions on
the phase space  ${\cal P}_{2n} \sim T^*\Sg(t) $, the cotangent  bundle
over $\Sg(t)$. They may be considered locally  as functions of  Darboux
coordinates $\xi^i, \ p_j $, with  $\{\xi^i\} \in \Sg(t)$ and
\beq
p_k  = \frac{mc\ \omega_{kl} (t, \xi) \,\dot \xi^l}{\sqrt{ c^2 -
\omega_{ij}(t, \xi) \,\dot \xi^i\,\dot \xi^j}}.
\nde
Of course, the Darboux coordinates fixed by a choice of coordinates
$\xi^i$ on $\Sg(t)$  are observables,too. They form the so--called
{\it primary observables} in the sense that other observables are
functions  of them.\\

{\bf 2.2.  GENERAL CONCEPT OF CANONICAL QUANTIZATION}\\

Consider now the concept  of quantization  of a classical Hamiltonian
system. A general definition of  the  asymptotical quantization
can be found in  the  book by Karassiov and Maslov (1991), Chapter IV.
We shall adopt the following  simplification of the deformational
version  of this  definition. (The simplification consists in that
the deformed Poisson brackets are supposed in condition (Q2) below
instead of the usual definition which starts with a $*$-product of symbols of
operators.)

Let $ {\bf s}_{2n}$ be  an appropriate subalgebra of the Poisson algebra
of functions $ f \in C^\infty({\cal P}_{2n})$.
{\it Quantization} is  a map \\

\beq
{\cal Q}: \ {\bf s}_{2n} \ni f
\qarr \hat f\quad (\mbox{operators in a Hilbert space}\ {\cal H}), \nonumber
\nde
satisfying the following conditions:

(Q1) $ \quad  1\ \qarr {\bf\hat 1} $\ (the identity operator in $\cal H$);

(Q2) $\quad  \{f, g\}_\h \ \qarr \
 \im\h^{-1} [\hat f, \hat g] \defst
\im\h^{-1}   (\hat f \hat g - \hat g \hat f)$ \quad
where $\{f, g\}_\h \equiv  \{f, g\}_0 + O(\h) $ is an antisymmetric
bilinear functional of $f$ and  $g$ and  $\{f, g\}_0 \equiv \{f, g\} $ is
the Poisson bracket  in ${\cal P}_{2n}$;

(Q3) $\quad \ \hat{\ov f} \qarr  (\hat f)^\dagger $
(the Hermitean conjugation of  $\hat f$ with respect to the scalar
product in $\cal H$);

(Q4)   a complete set of functions (maximal Abelian subalgebra)
 $f^{(1)}  , ..., f^{(n)}: \  f^{(i)} \in {\bf s}_{2n},$
is mapped to a complete set (in the sense by Dirac (1948),
Chapter.III) of commuting operators $\hat f^{(1)}, ..., \hat f^{(n)}$.\\

It follows also from the condition (Q4) and the
Stone--Von Neumann theorem that
${\cal H} \sim L^2 (\Sg; {\Bbb C}; \sqrt \om d^n x)$.

The main problems  of quantization consist in an  infinite number of
possibilities to construct the functional $\{f, g \}_\h$
(deformation of the Poisson bracket),  in difficulties with
construction of a complete set  on  the topologically non-trivial
spaces  ${\cal P}_{2n}$  and  in diffeononinvariance of quantum
observables.
Here we have a simple and physically oriented purpose to consider
traditional procedures of quantization  in application  to a particular
elementary system on a class of simple but non-trivial geometric backgrounds.
Therefore,  the following restrictions on  the system and spaces
\rin and  \rn under consideration will be supposed.

(V1) Assume that  \rin  is a  globally static space-time and
 $\Sg (t) \sim \rnf $ are its completely geodesic sections that exist in
this case. It means that $\om_{ij} (\xi, t) \equiv \om_{ij} (\xi) $.
Then, the classical dynamics' with the Hamilton functions  $H$ and $H_0$
are equivalent  and refer only  to different systems of reference.

(V2) Our main purpose is  to construct a quantum image  of  the
Hamilton function (classical Hamiltonian)  $H_0$ for an  arbitrary
$\om_{ij} (\xi) \in C^\infty (\rnf)$  starting  with the  general
 scheme of quantization (Q1)--(Q2). The minimal algebra ${\bf s}_{2n}$
containing all such Hamiltonians is   the algebra of  polynomials
in  $p_i$ with the coefficients  depending on  $\xi^i$.  If
a nonrelativistic quantum Hamiltonian $\hat H_0$ is  constructed,
then a possible way to obtain a relativistic one $\hat H$ is provided
by the Von Neumann rule (Von Neumann, 1955, p. 313)  defining functions
of commuting  operators  $\hat A_1, ... , \hat A_N$:
\beq
f(A_1,..., A_N ) \qarr   \hat f \defst f (\hat A_1, ... , \hat A_N).
\label{neu}
\nde
Being applied to  the classical Hamiltonian (\ref{ham}) interpreted  in
the asymptotical sense, it  gives
\beq
   {\hat H} (H_0) = H(\hat H_0 )
\defst  \hat H_0 - \frac1{mc^2}\hat H_0^2
+ \frac1{2m^2 c^4}\hat H_0^4 - ...    \label{hamn}
\nde

(V3) Assume that the topology of  \rn is  trivial;  of course, it does
not mean that the curvature  of \rn  along the metric  $\om_{ij}$ is zero.
The physical meaning of this condition  may not be considered as
a restriction  on  the topology  but  as  localization of the
quantum particle in a sufficiently small  domain
 so  that only local manifestations of the space curvature are essential.

(V4)  In  virtue of the preceding  assumption,
 it is  supposed   that  the coordinate lines
 $\xi^i $  on  \rn  are complete and open.  In this sense, they are similar
to the Cartesian coordinates.

By the way, under assumptions  (V1) -- (V4),
there are no QFT-process of
creation  and annihilation of  particles by the external gravitational
field, and  the quantum dynamics becomes a  purely quantum--mechanical one.

{\it The canonical quantization}  means here the following realization
of $\cal Q$:

(CQ1) One takes  some coordinates $\xi^i$ satisfying  (V4)   as  a complete
set $f^{(1)}, ..., f^{(n)}$ in the condition (Q4).

(CQ2) One takes,  at first,  the algebra of polynomials in the Darboux
coordinates $\xi^i, p_j $  as the  algebra $ {\bf s}_{2n}$.

(CQ3)  One imposes    the following conditions
on the  functional  $\{f, g\}_\h$:
\beqa
\{\xi^i, \xi^j\}_\h &=& \{\xi^i, \xi^j\}_0 \equiv
\{\xi^i, \xi^j\} = 0, \\
\{\xi^i, p_j\}_h &=& \{\xi^i, p_j\}_0 \equiv \{\xi^i, p_j\} = \delta^i_j, \\
\{p_i, p_j\}_\h &=& \{p_i, p_j \}_0 \equiv
\{p_i, p_j\} = 0,
\ndea
 thus making the condition (Q2) more definite.

(CQ4) Then,  the quantum images  $ \hat\xi^i, \hat p_j $ of these
primary classical  observables should satisfy  the canonical
commutation relations
\beq
[\hat\xi^i, \hat\xi^j] = 0,
\quad [\hat\xi^i, \hat p_j] = \im\h {\delta^i}_j,
\quad [\hat p_i, \hat p_j] = 0. \label{ccr}
\nde
and may be realized  as differential operators\footnotemark[3]
\beqa
   \xi^i \qarr    \hat\xi^i &=& \xi^i \cdot\hat {\bf 1},\nonumber \\
   p_i \qarr \hat p_j  &=& - \im \h(\ptl_j + \frac14 \ptl_j \ln \omega).
      \label{xp}
\ndea
in $ L^2 (\Sg (= \rnf); {\Bbb C}; \sqrt\omega d^n \xi)$

(CQ5) Further,  one maps the  basis of $ {\bf s}_{2n}$  formed by the unity
 and monomials
\beq
(\xi^1)^{M_1}...(\xi^n)^{M_n} (p_1)^{N_1}...(p_n)^{N_n}    \label{mn}
\nde
onto the identity operator ${\bf\hat 1}$ and hermitizations
of the same monomials formed by operators  $\hat\xi^i,\ \hat p_j$
along a  chosen {\it rule of ordering} ( an Hermitean arrangement of the
operators in each monomial).  As usual, the rule determines the functional
$\{f, g\}_\h$ in the  quantization  condition (Q2)  via   commutation
of the operators thus obtained, though it is not known to the present
author if any rule determines  the functional.

(CQ6) The functional $\{f,g\}_\h $ fixed by a rule of ordering
is taken further  as  the general relation  (Q2) for any
$f(\xi, p),\ g(\xi,p) \in C^\infty({\cal P}_{2n})$ in view of the
density of the polynomials  in   $C^\infty $, see Berezin and
Shubin (1984).

In general, there are infinitely many possible rules of ordering
and a classification of them, apparently not exhausting, is given
by Agarwal and Wolf (1970).

The {\it Weyl rule} (Weyl, 1931)  is the most popular one in the literature.
It has some attractive symmetry properties, see, for instance, Mehta (1964).
For the particular  case under consideration, it may be described as
follows. Consider, for example,  the following product
$({\hat p}_1)^a ({\hat\xi^1})^b ,  a \ge 0, b \ge 0, $ of
non-commuting operators. Then, the Weyl ordering
$\left(({\hat p_1})^a  ({\hat\xi^1})^b\right)^{({\rm W})} $
of the product is determined  by the following relation, Berezin
and Shubin (1984), Chapter 5:
\beq
\left(A{\hat p_1} +  B {\hat\xi^1}\right)^N =
\sum_{a+b = N} \frac{N!}{a!b!} A^a B^b
\left(({\hat p_1})^a   ({\hat\xi^1})^b\right)^{({\rm W})}.
\nde
If one takes the Weyl ordering, then the functional
$\{f, g\}_\h \defst \{f, g\}_\h^{({\rm W})} $ is the Moyal bracket
(Moyal, 1949). Taking into account the Riemannian measure on  $\Sg$
and condition (V3) on coordinates $\xi$,  one can represent the
canonical quantization
of  the polynomials $f(\xi, p)$ via the Weyl ordering as follows
(Berezin and Shubin (1984), Chapter 5):
\beqa
f(p, \xi)  &\qarr\ &  (\hat f^{({\rm W})} \psi) (\xi) \nonumber\\
&=& (2 \pi \h)^{-n} \omega^{-\frac14} (\xi)\,
\int d^n\xi'\,  \, d^n p\
\exp\left(-\frac{\im}{\h} (\xi^i - {\xi'}^i) p_i \right)
f\left(p, \frac{\xi +\xi'}{2}\right)\ \omega^{\frac14}(\xi')\,
\psi(\xi'), \label{wey} \\
&& \qquad\qquad
\psi(\xi)  \in   L^2 (\Sg ; {\Bbb C};  \sqrt \omega\, d^n \xi)
\nonumber
\ndea
Further, in view of the mentioned density of the polynomials in
$C^\infty ({\cal P}_{2n} \equiv R_n \bigotimes\rnf)$,
this correspondence is adopted as a general definition of
the canonical quantization of $f(p, \xi) \in C^\infty ({\cal P}_{2n})$.

Another example is {\it the Rivier rule of ordering} (Rivier, 1957,
Mehta, 1967)
which, in application to a monomial (\ref{mn}) is the following
arrangement of the primary observables:
\beqa
(\xi^1)^{M_1}...(\xi^n)^{M_n} (p_1)^{N_1}...(p_n)^{N_n} &\qarr&  \nonumber
\\
 \qarr \frac12\left((\hat\xi^1)^{M_1}...(\hat\xi^n)^{M_n} (\hat p_1)^{N_1}...
(\hat p_n)^{N_n}\right. &+&  \left.(\hat p_1)^{N_1}... (\hat p_n)^{N_n}
(\hat\xi^1)^{M_1}...(\hat\xi^n)^{M_n}\right) \nonumber\\
\defst   \left((\hat\xi^1)^{M_1}...(\hat\xi^n)^{M_n} (\hat p_1)^{N_1}...
(\hat p_n)^{N_n} \right)^{({\rm R})}. && \label{riv1}
\ndea
Similarly to the Weyl ordering, it can be represented in the form
\beqa
& &f(\xi, p)\qarr\ (\hat f^{({\rm R})} \psi) (\xi) \nonumber\\
& & = (2 \pi \h)^{-n}
\omega^{-\frac14} (\xi)  \int d^n\xi'\,  d^n p\
\exp\left(-\frac{\im}{\h} (\xi^i - {\xi'}^i)p_i\right)\
\frac{f(\xi, p)
+ f(\xi', p)}{2} \  \omega^{\frac14}(\xi') \, \psi(\xi'), \label{riv} \\
&& \qquad\qquad
\psi(\xi)  \in   L^2 (\rnf; {\Bbb C};  \sqrt \omega\ d^n \xi),
\nonumber
\ndea
which is obtained as the half--sum  of the integral representations of
$qp$- and $pq$-orderings given   by Berezin  and Shubin (1984), Chapter 5.
Again, the  rule  (\ref{riv}) is  extended  to all
$f(\xi, p) \in C^\infty ({\cal P}_{2n})$.
 To the Rivier ordering, its own "bracket" corresponds in the condition (Q2),
which may be denoted  as  $\{f, g\}_\h^{({\rm R})}$.

Rewrite (\ref{wey}) and   (\ref{riv}) in a  compact form
\beq
(\hat f^{({\rm W})} \psi) (\xi) =
\int d^n\xi' K_f^{({\rm W})} (\xi;\xi') \psi(\xi')\quad {\rm and} \quad
(\hat f^{({\rm R})} \psi) (\xi)
= \int d^n\xi' K_f^{({\rm R})} (\xi;\xi') \psi(\xi')
\nde
It is  obvious  that the kernels   of the form
\beq
K_f^{(\nu)}{({\rm W})} (\xi, \xi') \defst \nu K_f^ (\xi, \xi')
+ (1-\nu) K_f^{({\rm R})} (\xi, \xi')\quad  \label{alk}
\nde
define   an ordering for any fixed  value of the real  parameter $\nu$,  too,
and, in general, there are  many other possibilities of such linear
combinations. \\

{\bf 2.3. QUANTIZATION OF  GEODESIC MOTION}\\

It is  the time now  to return to the
concrete system  we  intend to quantize, namely, the system
described by the Hamilton function   $H_0 (\xi, p)$.
An important point is that  we apply the  Von Neumann rule (\ref{neu})
to  the metric tensor $\om_{ij} (\xi)$:
\beq
\om^{ij} (\xi) \qarr \hat\om \equiv \om^{ij} (\hat\xi)
= \om^{ij}({\xi}) \cdot {\bf \hat 1}.
\nde
 Suppose also that should not appear {\it operators} of the form
$$
\widehat {(\ptl_{i_1}...\ptl_{i_M} \om^{ij} (\xi))}, \quad M > 0,
$$
should not appear in the canonical  Hamilton operator $\hat H_0$
which  we are looking for.   It  does
not mean that the representation  of $\hat H_0$ as a differential
operator in
$ L^2 (\rnf;\ {\Bbb C};\  \sqrt \omega\ d^n \xi)$ should not contain
{\it functions} of the form $\ptl_{i_1}...\ptl_{i_M} \om_{ij} (\xi)$.
Then, it is easy to see that all possibilities to choose
 rules of ordering for quantization along the scheme (CQ1)--(CQ2)
are reduced to a  one--parametric family (\ref{alk}).
Simply speaking, the possible  orderings  are all  those Hermitean
arrangements of the operators  $\hat p_i $ ¨ $\hat \om^{jk}$, which
reproduce  the classical Hamiltonian $H_0$ under  assumption
that the operators commute and thus satisfy to the Correspondence Principle.
 However, if, for example,
a system with a classical Hamiltonian of the form
$\lambda^{ijkl} (\xi) \, p_i p_j p_k p_l $ were considered
the one-parametric family of  kernels (\ref{alk}) would not exhaust
all possible  orderings. The latter  would form apparently a
two--parametric family and, thus, the ambiguity became larger.

Quantization of $H_0$ along the rule  (\ref{alk})
gives the following  correspondence after use of  (\ref{wey}) and
(\ref{riv}):
\beq
 H_0 (q,p) \qarr \hat H_0^{(\nu)} =
\frac{2-\nu}{8m} \om^{ij} (\hat\xi) \hat p_i \hat p_j
+ \frac\nu{4m} \hat p_i  \om^{ij} (\hat\xi) \hat p_j
+  \frac{2-\nu}{8m}\hat p_i \hat p_j \om^{ij} (\hat\xi). \label{hal}
\nde
Substituting here representations (\ref{xp}) of the primary operators,
one obtains  $\hat H_0^{(\nu)}$ as a differential operator in
$ L^2 (\rnf;\ {\Bbb C};\  \sqrt \omega\ d^n \xi)$
\beq
  \hat H^{(\nu)}_0\ =
- \frac{\h^2}{2m} \Delta_{(\om)} (\xi)
+ V^{(\nu)}_{\rm q} (\xi),  \label{alh}
\nde
where  $\Delta_{(\om)}$ is  the Laplace--Beltrami operator for \rn,
\beq
V^{(\nu)}_{\rm q} (\xi) =
-\frac{\h^2}{4m}\left(\ptl_i(\om^{ij}\ga_j)
+ \frac{\nu}2 \ptl_i\ptl_j \om^{ij} +
 \frac {1-\nu}2 \om^{ij} \ga_i \ga_j\right) \label{val}\\
\nde
is the so--called {\it quantum potential} and
\beq
\ga_i \defst \ga^j_{ij}, \quad
\ga^k_{ij}  \defst \frac12  \om^{kl}
(\ptl_i \om_{jl}  + \ptl_j \om_{il} - \ptl_l\om_{ij})
\nde
are the  Christoffel symbols. Contrary to the kinetic term
$\hat H_0^{({\rm kin})} \defst  - (\h^2/2m) \Delta_{(\om)} (\xi) $,
the quantum potential is not diffeo-invariant. In the generic case,
there is no choice of
coordinates $\xi$ for which   $V^{(\nu)}_{\rm q}(\xi) \equiv 0 $
in a  domain; it is  easily seen from consideration of the
integrability condition of the  equation
$ V^{(\nu)}_{\rm q} = 0$. In this sense,  the quantum potential
distinguishes no preferred coordinate system.  This dependence of the
quantum dynamics on coordinate systems can be called
apparently {\it a quantum anomaly of diffeomorphisms of the configurational
space}.

Thus, the arbitrariness  in construction  of quantum mechanics
of a particle in \ri is contained in  the quantum
potential $ V^{(\nu)}_{\rm q}$  in the form of its dependence on
the parameter $\nu$  and on a choice of coordinates $\xi^i$.
This arbitrariness is not trivial because it leads to Hamilton operators
with different spectra. Some authors  eliminate it "by hand"
setting simply $\hat H_0 \equiv  \hat H_0^{({\rm kin})}$.  Just so
Gavrilov and Gitman (2001)  do  in fact. They consider
the space $ L^2 (\rnf;\ {\Bbb C};\   d^n \xi)$  and
 take  there as $\hat H_0$
(in their own notation)  the  operator
\beq
\hat H_0^{({\rm GG})}
= \frac1{2m} \hat\om^{-\frac14}  \hat p_i
\hat\om^{\frac12} \hat \om^{ij}
 \hat p_j \hat\om^{\frac14} \equiv   \label{kin}
- \frac{\h^2}{2m} \sqrt \om \Delta_{(\om)}
\nde
which is equivalent to $H_0^{({\rm kin})}$; here, of course,
$\hat\om^{ij} \equiv \om^{ij} (\xi) \cdot {\bf \hat 1}$.
 The correspondence principle is evidently satisfied: if
one assumes that   $\hat\xi$ and $\hat p$ commute then he comes to
$H_0$.  A problem, however, is   to go a way in the reverse direction
and to obtain the rule (\ref{kin}) as a Hamilton  operator
along a more or less well formulated quantization formalism
Representation (\ref{kin}) can be found in  the paper  by
DeWitt (1957) but namely as the kinetic part of the total Hamilton operator
which  includes also  a quantum potential.
A brief exposition  of this  result and its application for
elimination of the ambiguity of the canonical quantization described
by the parameter  $\nu$ will be given in the following chapter.\\

{\large\bf 3. QUASICLASSICAL QUANTIZATION OF GEODESIC\\ MOTION}\\

{\bf 3.1. DEWITT'S HAMILTONIAN AND RIEMANNIAN COORDINATES}\\

B.DeWitt (1957) generalized to \rn the WKB--propagator proposed by
Pauli (1950) for a particle in the electromagnetic field  in \eu.
As a result, the following nonrelativistic propagator  was  obtained:
\beq
<\xi,t| \xi_0, t_0> =
\omega^{-1/4} (\xi) D^{1/2} (\xi,t |\xi_0,t_0 )\omega^{-1/4} (\xi)\
\exp\left(-\frac\im\h S(\xi,t|\xi_0, t_0) \right), \label{prop}
\nde
where $ D $ is the Van Vleck determinant (Van Vleck, 1928 )
\beq
D (\xi,t |\xi_0,t_0 )  \defst
\det\left(- \frac{\ptl^2 S(\xi,t|\xi_0,t_0)}{ \ptl \xi^i \ptl \xi^j_0}
\right),
\nde
and
\beq
S(\xi, t|\xi_0, t_0)
= \int_{t_0}^t \frac12 \om_{ij} (\xi, t) \dot \xi^i \dot \xi^j, \qquad
\xi_0 \defst \xi(t_0)  \label{act}
\nde
is the classical action; its minimum is provided by the following
equation of motion.
\beq
\ddot{\xi}^i  + \ga^i_{kl}(\xi; t)  \dot{\xi}^k \dot\xi^l +
\om^{ik} (\xi; t) \frac{\ptl\om_{kl}}{\ptl t} (\xi; t)
\dot\xi^l = 0 \label{geo}
\nde
If  $\ptl\om_{kl} (\xi; t)/\ptl t = 0 $, that is,  if  \rin is the
globally static space-time  and   $\Sg(t)\sim \Sg(t_0)\sim \rnf $
is  a completely geodesic hypersurface,  then  (\ref{geo})
is the geodesic equation in \rn. Restrict our consideration to
this simple case,  the more so  that  DeWitt does, in fact, the same.

 Considering the limit $t \rightarrow t_0 \ (\xi\rightarrow \xi_0) $
{\it along the geodesic line, connecting $\xi$ and $\xi_0$}, DeWitt
comes to the equation
\beq
\im \h \frac\ptl{\ptl t}<\xi|\xi_0> +
 \frac{\h^2}{2m} \left(\Delta_{(\om)} (\xi)
- \frac16 R_{(\om)} (\xi)\right) <\xi|\xi_0>
= o (\xi - \xi_0)) <\xi|\xi_0>,
\label{DW2}
\nde
where $R_{(\om)}$ is  the scalar curvature for the metric $\om_{ij}$;
the Riemann-Christoffel and Ricci tensors being defined as follows:
\beq
   {R^a}_{bcd} =\ptl_d {\gamma^a}_{bc} - \ptl_c {\gamma^a}_{bd} +
  {\gamma^a}_{de}{\gamma^e}_{bc} - {\gamma^a}_{ce} {\gamma^e}_{bd} \quad,
\qquad R_{(\om) ij} = R^k_{(\om)ikj}. \label{RK}
\nde
(DeWitt's definition of  $R_{(\om) ij}$ has  an opposite sign.)
It follows from  (\ref{DW2}) that the differential operators
\beq
\hat H^{({\rm DW})}_0 (\xi) =
 -\frac{\h^2}{2m} \left(\Delta_\om (\xi)
- \frac16 R_{(\om)} (\xi)\right), \label{hamR}
\nde
can be considered as the Hamilton operator on the subspace of the wave
functions (initial data for the \Sche)
$$
<\xi|\xi_0>\equiv \psi_{\xi_0} (\xi) \in
L^2 (\rnf; {\Bbb C};  \sqrt \omega d^n \xi) $$
 which  are localized
in  a  small neighborhood of the point $\xi_0$  in the sense that they
satisfy the condition
\beq
\frac{(\psi, o(\xi -\xi_0)\psi)}{(\psi, \hat H^{({\rm DW})}_0  \psi)} \ll 1,
\nde
where  $ o(\xi -\xi_0)$ is a residual term in the right-hand side of
(\ref{DW2}).  Thus,  the approach exposed which is relevant to call
the quasi-classical one gives, in the mentioned {\it  approximate} sense,
a unique  and  diffeoinvariant  Hamilton operator.

At the same time, DeWitt and  some other  authors considered the
appearance of the potential  $  (\h^2/12m) R_{(\om)} (\xi) $ as an
unfavorable phenomenon because traditionally  $\hat H_0^{({\rm kin})}$
was taken as the Hamiltonian of the particle in \rn. They added
an appropriate counterterm  into the  Lagrangian,
that is into the integrand in formula (\ref{act}), to have
$\hat H_0^{({\rm kin})}$ instead of $\hat H^{({\rm DW})}_0$
in eq.(\ref{DW2})/
 However, the corrected Lagrangian  is not  the one  of  geodesic
motion of which quantization is the matter of the present
paper.  Actually, the appearance of  $R_{(\om)}$ in the Hamiltonian
is  quite in conformity with the QFT-approach
 (Tagirov, 1999)  a brief exposition of which will be given
in Section 6.\\

{\bf 3.2. COMPARISON OF  CANONICAL AND DEWITT'S HAMILTONIANS }\\

Now,  let us compare the Hamiltonian  $\hat H^{(\nu)}_0$, obtained
exactly  in the  canonical sense  and  the  approximate
quasiclassical  one  $\hat H^{({\rm DW})}_0$.  Remind  that the
 latter was obtained  by retracting the point  $\xi$ to  $\xi_0$
{\it along a geodesic line} connecting them. Thus, a position of
$\xi$ with  respect  to  $\xi_0$ naturally defined  by the geodesic
distance  $s(\xi,\,\xi_0)$  between them and the tangent vector
$(d\xi^i/ds)_0 $ to the geodesic line at $\xi_0$. These quantities form
{\it the  Riemannian coordinates}
\beq
y^i (\xi) \defst  s(\xi,\,\xi_0)\left(\frac{d\xi^i}{ds}\right)_0 \label{y}
\nde
with  the origin at the point $\xi_0$. In these coordinates the metric
tensor $\om_{ij}$, its derivatives  by  $ y^i$ and, respectively,
the Christoffel symbols $\ga^i_{kl}$  are  represented as  a power series
in $ y^i$,   coefficients of which are polynomials in powers of
the components of the Riemann--Christoffel tensor and of its covariant
derivatives taken at the origin  $ y^i = 0 $, i.e.,  at  $\xi_0$.
Therefore, applying the Veblen method of affine extensions
(Veblen, 1928) using contracted Bianchi identities,  one can represent
the quantum potential $V_{\rm q}^{(\nu)}$ as a similar series. For our
discussion, the following two terms of the series are sufficient:
\beq
{\hat H}^{(\nu)} _0 (y)  =  -\frac{\h^2}{2m} \left(\Delta_{(\om)} (y)
-\frac{\nu}{12}\left. R_{(\om)} \right|_{y=0}
- \frac{\nu}{12} \left.(\ptl_i R_{(\om)})\right|_{y=0}\ y^i +
O((y)^2)\right).  \label{hamy}
\nde
The condition of coincidence of ${\hat H}^{(\nu)} _0 (y)$ with
$\hat H^{({\rm DW})}_0$  in the zero order approximation is satisfied
for the value  $ \nu = 2 $   in  (\ref{alk})
(\ref{hal}). Thus, from the canonical  point of view adopted here,
{\it the correct nonrelativistic Hamilton operator} for  a point-like
particle in the globally static  \rin  is the following
remarkably simple  expression:
\beq
 \hat H_0^{(2)} = \frac1{2m}\hat p_i \om^{ij}(\hat \xi) \hat p_j .
\nde
This solves  the ambiguity  problem  of  ordering  of the primary
operators  in the canonical quantization of the geodesic motion.
However,  the  problem of diffenoninvariance of quantum potential
$V_{\rm q}^{(2)}$  retains. This
problem, as well as  the problem
of ordering for the Hamiltonians which are not quadratic in momenta
will be  discussed in Section 5. And  now we pass to a justification
of the obtained result coming from
consideration of another traditional approach to formulation of
quantum mechanics.\\

{\large \bf 4. QUANTIZATION OF GE0DESIC MOTION BY  PATH \\
INTEGRATION}\\

{\bf 4.1. RELATION BETWEEN CANONICAL AND PATH INTEGRATION\\ FORMALISMS}\\

Not only the point-like particle motion but also a number of other
mechanical problems  are  naturally represented  as  a  geodesic
motion or its  generalization in  some \rn. Usually, the latter are
homogeneous spaces of  symmetry groups, see, for instance, Marinov (1995),
 Groshe, Pogosyan  and Sissakian (1997) and references therein.
For this class of systems, the Feynman formalism
of path integrals Feynman (1949, 1951) is considered as a
very appropriate approach
to solve the \Sche for the particle propagator  since it takes into
account the metric of the configurational space through a natural measure
and representation of the vitrtual path as consisting of small segments
of geodesic lines.

In  this approach, the path integral relates a given quantum
Hamiltonian $\hat H_0$  represented as  a differential operator in
$L^2 (\rnf; {\Bbb C};  \sqrt \omega d^n \xi) $ to some  effective
classical Lagrangian (Marinov, 1995). The Hamiltonian may be considered as
a result of quantization of the classical dynamics described by the
Lagrangian so found. An inverse problem can be posed: to select
$\hat H_0 \defst \hat H_0^{(\rm F)}$ (the superscript $({\rm F})$ denotes
"Feynman" as will be clear a bit below)   so that the effective
Lagrangian would prove to be  the classical one for the geodesic motion:
\beq
L_{\rm eff} (\xi, \, \dot\xi) = L_{\rm cl} (\xi, \, \dot\xi)
\equiv \frac m2 \om_{ij}(\xi) \dot \xi^i \dot \xi^j, \label{leff}
\nde
A correspondence   $ H_0 \rightarrow \hat H_0^{(\rm F)} $ thus defined
and taken together with the map  (\ref{xp}) of  the primary  observables
may be  called  {\it the Feynman quantization}
of the geodesic motion in  \rn.
Consider such an approach in  a brief descriptive form sufficient for
a comparison with the formalisms considered above.

So,  a problem is to represent, as a path integral, the following
formal propagator in \rn
\beq
{\Bbb K} (\xi'', t''|\xi', t')
= <\xi''| e^{-\frac\im\h (t''- t')\hat H_0}| \xi'>, \label{K1}
\nde
for the quantum Hamiltonian of the form
\beq
   \hat H_0 = - \frac{\h^2}{2m} \Delta_\om (\xi)  +  V(\xi).  \label{HAMV}
\nde
acting in $ L^2 (V_n;\, {\Bbb C};\,  \sqrt \omega\, d^n \xi)$.
Here we consider as an  already  established fact that the  set
of the  possible (non-relativistic)  Feynman Hamiltonians
$ \hat H^{(\rm F)}_0 $  a particle in \rn  is  contained
among  Hamiltonians (\ref{HAMV}) with arbitrary potentials $V(\xi)$.

Following the line of calculations  by D'Olivo and Torres (1989),
divide the time interval  $[t',\, t'']$ by  $N \rightarrow\infty$
intervals  of infinitesimal duration  $\epsilon = (t''- t')/N $ and
represent  $ {\Bbb K} (\xi'', t''|\xi', t') $ as follows:
\beq
{\Bbb K} (\xi'', t''|\xi', t') = \lim_{N\rightarrow\infty} \int \
\prod_{I=1}^{N-1} \sqrt{\omega (\xi_I)}\ d^n\xi_I \prod_{J=1}^{N-1}
<\xi_I|e^{-\frac\im\h \epsilon\hat H_0}|\xi_J>,     \label{K2}
\nde
where $\xi_0 =\xi' , \ \xi_N =\xi''$.

To calculate  the matrix elements
of $\hat H_0 $  in the configurational representation  one should
represent the  differential operator  $\Delta_\om (\xi)$
in  (\ref{HAMV}) through  $\hat \xi, \hat p$. To this end, {\it a  rule
of ordering of them} should be fixed.   Contrary to
D'Olivo and Torres (1989),
who, as  many other authors on the matter, adopted  the Weyl rule,
we use a more general rule (\ref{alk}).
Then,  we have
\beq
   \hat H_0 =  \hat H^{(\nu)}_0 - V_{\rm q}^{(\nu)} (\xi) + V(\xi), \label{h}
\nde
where $\hat H^{(\nu)}_0$ and   $V_{\rm q}^{(\nu)} (\xi) $ are assumed
to be   expressions (\ref{hal}) and (\ref{val}) respectively.
Calculation  of the matrix elements within the  terms  linear in $\ep$
using our generalized rule of ordering gives:
\beqa
{\Bbb K} (\xi'', t''|\xi', t') &=& \lim_{N\rightarrow\infty} \int \
\left(\frac{1}{2\pi \im\h\epsilon}\right)^{\pi N/2}
 \prod_{I=1}^{N-1} \sqrt{\om (\xi_I)}\ d^n \xi_I \nonumber\\
&\times&\prod_{J=1}^{N-1}
\frac{\left(\tilde{\sqrt\om}\right)^{(\nu)}
(\xi_{J-1},\,\xi_J)}{[\om(\xi_J) \om(\xi_{J-1})]^{1/4}}
\exp\left\{\frac{\im}{\h} \epsilon
{\tilde L}^{(\nu)}_{\rm eff}\left(\xi_{J-1}, \xi_J;
\frac{\Delta \xi_J}{\epsilon}\right)\right\}, \label{K3}\\
\Delta\xi_J &\equiv&  \{\Delta\xi^i_J \defst
\xi^i_J - \xi^i_{J-1}\}.\nonumber
\ndea
Here   $(\tilde {\sqrt\om})^{(\nu)} (\xi_{J-1}, \xi_J)  $ ¨
${\tilde L}^{(\nu)}_{\rm{eff}}\left(\xi_{J-1}, \xi_J,
\Delta \xi_J/\epsilon \right)$
are  the quantities that are expressed, respectively, through the functions
${\sqrt\om (\xi)} $ and
\beq
L_{\rm{eff}}^{(\nu)}\left(\xi, \frac{\Delta \xi_J}{\epsilon} \right)
\defst L_{\rm cl} \left(\xi,\ \Delta \xi_J/\epsilon \right) -
 V(\xi)  + V_{\rm q}^{(\nu)}
\nde
along the following general rule implied by eq.(\ref{alk}):
\beq
{\tilde f}^{(\nu)}\left(\xi_{J-1}, \xi_J\right)
=  \nu  f(\bar\xi_J) + \frac{1-\nu}2 \left(f(\xi_{J-1}) + f(\xi_J) \right),
\quad \bar \xi_J \defst \frac 12 (\xi_J + \xi_{J-1})
\label{tilde}
\nde

Now, the product in $J$  in eq.(\ref{K3}) should be  represented  as
a product of exponentials of some  classical action on the  intervals
$[\xi_{J-1},\, \xi_J]$, that is  as  a product of factors  of the form
\beq
\exp\left\{\frac{\im}{\h} \epsilon
 L'_{\rm eff}\left(\xi'_J,\ \Delta \xi_J/\epsilon\right)\right\},
\label{L'}
\nde
where, in the exponent, the value of some effective Lagrangian
$ L'_{\rm eff}(\xi, \dot\xi) $  (in general, it differs from
$L_{\rm eff}^{(\nu)}$ )   stands,  which is taken at the   point
$ \xi'_J \in [\xi_{J-1},\, \xi_J]$  remained arbitrary for a
time being.

To obtain  the representation, all functions of
$\xi_{J-1},\ \xi_J, \bar\xi_J$  under the product in $J$
should be expanded into the Tailor series near the  point
$\xi'_J$ up to terms quadratic in  $\Delta \xi_J$,
since only such terms contribute to the integral eq.(\ref{K3}).
Further, one should include the contribution of the pre-exponential
factor to the exponent in a form of an additional quantum potential.
Consider this procedure separately for the two  principally
different cases\\

A) The intermediate point evaluation of the  integrands:
$\xi'_J = (1-\mu) \xi_{J-1}  + \mu \xi_J, \quad
0< \mu < 1 $,  i.e.,  $\xi'_J \in (\xi_{J-1},\, \xi_J)$.\\

B) The  end point evaluation of  the integrands:  $\xi' = \xi_{J-1}$ or
$\xi' = \xi_J, $ i.e., $\xi'$  is taken at the ends of the closed interval
$[\xi_{J-1},\, \xi_J]$.\\

{\bf 4.2. QUANTUM POTENTIAL FOR THE  INTERMEDIATE POINT \\
EVALUATION OF  INTEGRANDS (CASE A)}\\

For the  generic function  (\ref{tilde}),  one has
\beq
{\tilde f}^{(\nu)}\left(\xi_{J-1}, \xi_J  \right) = f(\xi'_J)  +
(\frac 12 - \mu)\ptl_i f(\xi'_J) \Delta \xi_J^i
+ \frac 12 (\frac{2-\nu}4 -\mu + \mu^2 )\ptl_i\ptl_j f(\xi'_J)
\Delta\xi_J^i\Delta\xi_J^j. \label{f}
\nde
Apply this general formula to
$f(\xi) \equiv  L_{\rm{eff}}^{(\nu)}(\xi, \Dx/\ep)$.
The last term in eq.(\ref{f}) turns out to be equal to
\beq
\frac 12 (\frac{2-\nu}4 -\mu + \mu^2 )
\ptl_i\ptl_j \om_{kl} (\xi'_J) \Delta\xi_J^i\Delta\xi_J^j
\frac{\Delta\xi_J^l}{\epsilon} \frac{\Delta\xi_J^k}{\epsilon}\
\label{ptl}
\nde
in the necessary order of $\epsilon$.

Further, we  use the result by  McLaughlin and Schulman (1971)
according to which  the following substitution can be  made under
the integration in  eq.(\ref{K3}):
\beq
    \Delta\xi_J^i \Delta\xi_J^j \rightarrow \im \epsilon\, \frac{\h}m
\om^{ij}(\xi'_J).                    \label{ch}
\nde
After this substitution in eq.(\ref{ptl}) and symmetrization of the
resulting expression  in indexes $i, j, k, l$, one comes to the quantum
potential
\beq
V_L^{(\nu; \mu)} (\xi'_J; \nu; \mu) = -\, \frac{ \h^2 }{12 m}
 (\frac{2-\nu}4 -\mu + \mu^2 )\,
\bigl(\ptl_i\ptl_j \om_{kl} (\om^{ij}\om^{kl}
+ 2 \om^{ik}\om^{jl})\bigr)(\xi'_J)   \label{VOm}
\nde
in addition to
$L_{\rm eff}^{(\nu)}\left(\xi'_J, \Delta \xi_J/\epsilon \right)$.
Another additional term here
\beq
\im\, (\frac 12 - \mu)\, \frac\h3 \ptl_i \om_{kl}(\xi'_J)
\left(\om^{kl}(\xi'_J) \frac{\Delta\xi_J^i}\epsilon
+ 2\om^{ik}(\xi'_J) \frac{\Delta\xi_J^l}\epsilon\right),
\nde
comes  from  the second term in eq.(\ref{f})  after the use of the same
substitution (\ref{ch}). It adds to  $L_{\rm eff}^{(\nu)}$ a term
which is proportional to  $\Dx^i/\ep \sim \dot\xi$ that is linear
in the velocity.  There is no such term in  $L_{\rm cl}$ and  there
is nothing to compensate it so that the condition (\ref{leff}) were

satisfied. Indeed,  the logarithm  of the pre-exponential factor
\beq
\tilde \Omega_J  = \frac{\left(\tilde{\sqrt\om}\right)^{(\nu)}
(\xi_{J-1},\,\xi_J)}{[\om(\xi_J) \om(\xi_{J-1})]^{1/4}} \label{W0}
\nde
does not contain  a term linear  in    $\ep \dot \xi $:
when  $\epsilon \rightarrow 0$:
\beqa
\tilde \Omega_J   & = & 1 -
\left(\frac{\nu}8 \ptl_i \ptl_j \ln\om (\xi'_J)
- \left(\frac{3 - \nu}{32}
-\frac{\mu}4 + \frac{\mu^2}4 \right)\,
\ptl_i\ln\om(\xi'_J) \, \ptl_j \ln\om(\xi'_J) \right)
\Delta\xi^i\Delta\xi^j \nonumber \\
&+& O\left((\Delta\xi)^2\right)  \equiv \Omega (\xi'_J, \nu; \mu).  \label{W}
\ndea
Therefore, to avoid  appearance of a term proportional to the velocity
in $L_{\rm eff}$ one should take
\beq
\mu = \frac 12,       \label{be}
\nde
 i.e.,  $\xi'_J = \bar \xi_J$, as it is taken  by  D'Olivo and Torres (1989)
who adopt the Weyl ordering (formula  (\ref{tilde})  for  $\nu =1$)
from the beginning.

 Taking into account the condition (\ref{be}) and the substitution (\ref{ch})
one can reduce the contribution of  $\tilde \Omega$  into the path  integral
(\ref{K3})  to that   one  more  quantum potential  $V_\Omega^{(\nu)}$
is  added to  $L_{\rm eff}^{(\nu)}$  in the exponent of the exponential:
\beq
V_\Omega^{(\nu)} (\xi)  = -\, \frac{\h^2}{2m} \om^{ij} (\xi)
\left(\frac \nu 8 \ptl_i \ptl_j \ln\om (\xi) -
\left(\frac{1 - \nu}{32} \right)\,
\ptl_i\ln\om (\xi)\, \ptl_j \ln\om (\xi)\right) +  O(\ep^2).
\nde
As a result, if one chooses in the initial formula (\ref{HAMV})
\beqa
V(\xi)  &\equiv&   V_{\rm A}^{({\rm F};\nu)} (\xi) \defst V_{\rm q}(\xi) +
V_L^{(\nu)} (\xi) + V_\Omega^{(\nu)} (\xi)\nonumber\\
&=& - \frac{\h^2}{24m} \left(\frac{2\nu+7}{2} \om^{ij}\om^{kl} -
  (5-2\nu) \om^{ik}\om^{jl}\right) \ptl_i \ptl_j \om_{kl} \label{Vf}
 \\
&+& \frac{\h^2}{4m} \left( \frac{\nu+2}{4} \om^{km}\om^{ln}\om^{ij} -
\frac{\nu-2}{4} \om^{im}\om^{jn}\om^{kl}- (\nu-2) \om^{im}\om^{kn} \om^{jl}
\right) \ptl_i\om_{mn}\ptl_j\om_{kl}, \nonumber
\ndea
then, in the integrand of the path integral, only the following
product  remains  in  the  required approximation
\beq
\prod_{J=1}^{N-1}\exp\{\frac{\im \ep}\h L_{\rm cl}(\bar\xi_J)\},  \label{S}
\nde
that is  a product of exponentials of the ratio of
the classical action  of the geodesic motion between the points
$\xi_{J-1}$ and  $ \xi_J$ to the Planck  constant  $\h$.

Thus, we have determined  a map $ H_0 \rightarrow  H_{0A}^{({\rm F};\nu)} $
of the Hamilton function of the geodesic motion in  \rn on operator
(\ref{HAMV}) with  quantum potential (\ref{Vf}) is a version
acting on  $L^2 (V_n;\, {\Bbb C};\, \sqrt \om d^n\xi)$ is a version
of the Feynman quantization of the geodesic motion in  \rn.
It is not diffeoinvariant as well
as   contains  freedom in  the choice of the value of parameter  $\nu$
corresponding to arbitrariness of the  ordering rule in the
canonical quantization.  Could one select $\nu$ so that
$ V_{\rm A}^{({\rm F};\nu)}(\xi) $ would  coincide with the result of the
quasi-classical quantization (\ref{hamR}) in the region where
such comparison is relevant, i.e., in a neighborhood of the origin
of the normal Riemannian coordinates  ${y^i}$?  The answer is no,
it is not possible because
\beq
V_{\rm A}^{({\rm F};\nu)}(y)  =  \frac{\h^2}{2m} \, \frac R3 (0) + O (y)
\label{F1}
\nde
{\it independently of the value of} $\nu$ and, actually,  independently
on the choice of $\mu$. Thus, the initial ambiguity of the
canonical quantization  not only retains but also  become larger
in the considered version of the Feynaman quantization.\\

{\bf 4.3. QUANTUM POTENTIAL FOR THE END--POINT EVALUATION OF
 INTEGRANDS (CASE B)}\\

  In this case, if one takes   $\mu = 0$ or $\mu = 1$,
again the inadmissible addition of a term linear in $\dot\xi$ to
the exponent of the exponential occurs. It is a consequence of
an asymmetric contribution of the endpoints of the interval
$[\xi_{J-1},\,  \xi_J]$ while, for a given function $f(\xi)$,
expression (\ref{tilde}) for
$\tilde f^{(\nu)}(\xi_{J-1},\,  \xi_J) $ depends on the endpoints
symmetrically.  However,
the following symmetric expression for $\tilde f^{(\nu)}$,
\beqa
\tilde f^{(\nu)}(\xi_{J-1},\, \xi_J)
&=&  \frac 12 f(\xi_{J-1}) +  \frac 12 f(\xi_J)
+ \frac{\nu}8 \bigl(\ptl_i f(\xi_{J-1})
- \ptl_i f(\xi_J)\bigr) \Delta\xi_J^i \nonumber\\
&+ &  \frac{\nu}{16}
\bigl(\ptl_i \ptl_j f(\xi_{J-1}) + \ptl_i \ptl_j f(\xi_J)\bigr)
\Delta\xi_J^i\Delta\xi_J^j + O \left((\Delta\xi_J)^3\right), \label{fb}
\ndea
can be easily obtained if $\xi_{J-1}$ and  $\xi_J$ are at a short
distance.  Applying this formula  to
$\tilde f^{(\nu)} \equiv {\tilde L}^{(\nu)}_{\rm eff}$
in the exponent in formula  (\ref{K3}),   one
should  consider  contributions of the adjacent intervals
$[\xi_{J-2},\,  \xi_{J-1}]$ and $[\xi_J,\,  \xi_{J+1}] $ at the points
$\xi_{J-1}$ and $ \xi_J$, respectively.

  The total contribution to the phase  at  $\xi_J$ of the terms of
${\tilde L}^{(\nu)}_{\rm eff}$, which are
linear in  $\Delta\xi$, is
\beq
i \frac {\nu\h}{8m} \ep \om^{kl}(\xi_J)\ptl_i \om_{kl}(\xi_J) \ddot\xi.
\nde
and it  can be neglected in the path integration.
Here,  substitution  (\ref{ch}) and  relation
\beq
 \Delta\xi_{J-1} = \Delta\xi_J - \ep^2 \ddot \xi_J + O (\ep^3)
\label{ddot}
\nde
are used.
Making these substitutions in the terms which are quadratic in
$\Delta\xi$  one  obtains that a quantum potential
\beq
V_{\rm B}^{(\nu)}  = - \frac{\nu\h^2}{24m}\,\bigl((\om^{ij}\om^{kl}
+ 2 \om^{ik}\om^{jl})\, \ptl_i \ptl_j \om_{kl} \bigr)  \label{VL}
\nde

is added to  ${\tilde L}^{(\nu)}_{\rm eff}$.

The contribution to the phase of the adjacent pre-exponential terms
 $\tilde \Omega_J$ and  $\tilde \Omega_{J+1}$,
\beq
\tilde \Omega_J \cdot\tilde \Omega_{J+1}
= exp(\ln\tilde \Omega_J + \ln\tilde \Omega_{J+1}),
\nde
can be calculated in a similar way.  To this end,  expand the terms in
the exponents in powers of   $\Delta\xi_J$  and  $\Delta\xi_{J+1}$
up to  $ O\left((\Delta\xi)^3\right) $ and collect the terms  with
the coefficients that depend on  $\xi_J$. The remaining  terms go over
to the analogous contributions at the  points $\xi_{J-1}$ and  $\xi_{J+1}$.
Then, using  relation (\ref{ddot}), one obtains the following function
of  $\xi_J$:
\beq
\frac{\nu-2}8 \,  \ep^2 \ddot\xi^i_{J+1} \ptl_i \ln\om
+ \left(\frac{2-\nu}{16} \ptl_i\ptl_j\ln\om
+ \frac 1{32} \ptl_i \ln\om \ptl_j\ln\om\right)\cdot
\Delta\xi^i_J \Delta\xi^j_J + O\left(\ep^2 \Delta\xi\right).
\nde
Obviously, the first term here can be neglected under the integration.
Hence,  using  substitution (\ref{ch}), one finds  a contribution
to the phase  at  the  point  $\xi_J$  in a form of the following
quantum potential:
\beq
V^{(\nu)}_\Omega  =  - \frac{\h^2}{32 m}\, \om^{ij}
\left( 2(2-\nu) \ptl_i\ptl_j\ln\om
+   \ptl_i\ln\om \ptl_j\ln\om  \right). \label{VW}
\nde
Then,  one should  put
\beq
  V(\xi) \equiv V_{\rm B}^{({\rm F};\nu)}(\xi) \defst V_{\rm q}^{(\nu)} (\xi) +
V_L (\xi) + V_\Omega^{(\nu)} (\xi)
\nde
in  (\ref{HAMV}) in  order to retain in the phase  only a ratio of
the classical action near the point $\xi_J$ for the time  interval $\ep$
to the Plank constant $\h $.

Now, let us consider  $V_{\rm B}^{({\rm F};\nu)} $ at the origin of the normal
Riemannian  coordinates   $y^i$.   Note  at once that
$ V_L^{({\rm F};\nu)} (y) = O (y) $  since
\beq
\ptl_i \ptl_j \om_{kl} (y) = \frac13 (R_{(\om)ikjl} + R_{(\om)iljk})(0)
+ O(y), \label{syng}
\nde
 see, for instance, Synge (1960).   A non-vanishing  contribution
into  $V_\Omega^{(\nu)}$  can be given only  by the first  term
in (\ref{VW}). The contribution vanishes identically if and only if
\beq
       \nu = \ 2.        \label{a2}
\nde

Thus, we come to a  remarkable justification  of the ordering rule
  which had been found   by  comparison of the canonical
and quasi-classical Hamiltonians in  Section 3.  At the same time,
we have fixed a unique  way to   calculate  the path integral and,
in particular,
a prescription to evaluate the integrand functions: they should
be evaluated at the nodes of the lattice of integration.  The
prescription   differs  from that  induced by  the Weyl ordering
according to  which the evaluation should be done in the mid-points  of
intervals of the lattice.

It should be  noticed also that the ordering corresponding to
$\nu = 2 $ was mentioned  among many other ones by
D'Olivo and Torres (1989),
but we have singled out it from a two-parametric (in  $\nu$ and $\mu$)
set of possible orderings with a necessity.
In the next section, a  question will be  discussed in particular why
the comparison of quantum Hamiltonians in a vicinity of the origin
of the Riemannian coordinates has a special geometric meaning. As for now,
we give the complete  expression for $ V_{\rm B}^{(2)}$:
\beq
V_{\rm B}^{({\rm F};2)}  = -  \frac{\h^2}{12 m} (2 \om^{ij}\om^{kl}
+\om^{ik}\om^{jl})\ptl_i\ptl_j \om_{kl} -
\frac{\h^2}{16 m}(2 \ptl_i\om^{ij}\ptl_j \ln\om
+ \om^{ij} \ptl_i \ln\om\,  \ptl_j \ln\om ). \\
\nde
Of course, this Feynman quantum potential  differs, in general, from
the canonical one:
\beq
V^{(2)}_{\rm q} (\xi) =
-\frac{\h^2}{4m}\left(\ptl_i(\om^{ij}\ga_j)
+  \ptl_i\ptl_j \om^{ij}
- \frac {1}2 \om^{ij} \ga_i \ga_j\right),
\nde
(i.e., eq.(\ref{val}) for  $\nu = 2$) and  the question remains,
which of the potentials  is "more correct". \\

{\large\bf 5. DISCUSSION OF THE RESULTS OBTAINED}\\

Thus, taking  $\nu = 2$ in eq.(\ref{alk}) is  proposed as a concrete
and unambiguous solution  of  the problem  of  arbitrariness
  the  ordering rule, one of the main difficulties of
the canonical quantum mechanics in \rn. However, the rule  is
obtained namely for observables
(Hamiltonians) which are quadratic  in momenta.
If one  attempts to adopt the logic of our
construction  for an observable of a  more complicated structure,
the rule of ordering thus  obtained  will determine its own  "bracket"
in the  quantization condition (Q2). It is unclear,  will this rule
be  unambiguous but, in any case, we come to the conclusion that
{\it for different classes of  observables, there should be used different
"brackets"}  $\{.,.\}_\h$  in condition (Q2). This conclusion may seem
rather  strange, but, at least, it does not contradict to the
known experimental data  since the  corrections to the Poisson bracket

in the left-hand side of condition (Q2) are very small and,
correspondingly, differences  of corrections for different
versions are small too.

Further, the  result    refers  to the nonrelativistic
version of the geodesic dynamics.
A more  difficult problem of quantization of the relativistic
version  remains,  which is based on Hamiltonian
 $H (\xi, p)$, eq.  (\ref{ham}).   Its  possible asymptotic solution
by the use of  Von Neumann's rule in the form of (\ref{hamn}) has already
been in Section 2. However, if for  any classical
Hamiltonian,  its own canonical quantization  has to be constructed,
 then,   the way which was followed for the Hamiltonian (\ref{ham0})
should be passed anew for  (\ref{ham}).
In this case, an analog of the  quasi-classical
Hamiltonian  $\hat H_0^{({\rm DW})}$ should apparently be the
quantum Hamiltonian  calculated  in the Blattner--Costant--Souriau
formalism for the terms of the asymptotic expansion (\ref{ham}).
It is calculated for the first four terms  by  Kalinin (2000) and  differs
from the result of an immediate  application of the  Von Neumann
rule (\ref{hamn}). An analysis of this difference  seems to be
an interesting task  for understanding   relations between
different formalisms of  quantization.

Let us pass now  to the important point  that, to determine
the rule    $\nu =2$,  it was principal to compare the Hamiltonians
 $\hat H_0^{(\nu)}, \hat H_0^{({\rm DW})}$,
$\hat H_{0\rm A}^{({\rm F};\nu)}$
and  $\hat H_{0{\rm B}}^{({\rm F};\nu)}$ in a vicinity of the origin
of the Riemannian
coordinates $y^a$    namely. Why this system  is  distinguished  among
all possible systems? An  answer is apparently
as follows. The  position of
a point   $ \{\xi^i\} $ is defined in the Riemannian system completely
by the geodesic line connecting the point with the origin
$\{\xi^i_0\}$ and, therefore, only by the metric of \rn.
Indeed,  according to eq.(\ref{y})  the normal
Riemannian coordinates $y^a (\xi) =  n^{(a)}(\xi) s(\xi;\, \xi_0)$ of
the point $ \{\xi^i\} $
are completely  defined by values of the geodesic distance
 $ s(\xi;\, \xi_0) $   and
projections  $n^{(a)} (\xi) = e_i^{(a)} \, (d\xi^i/ds)_0$
of the tangent vector to the geodesic line connecting $\xi$ and $\xi_0$.
The coordinate lines
\beq
y^1 = const,...,\, y^{k-1} = const, \, y^{k+1} = const, ...,\,
y^n = const,\quad 1\geq k \leq n,
\nde
are distinguished  by  that their  all $n$ curvatures vanish.
Imagine that a similar system of coordinates
realized  not by the geodesics  but by the lines determined by some
other equation. Take,  for example, the geodesic equation  with an
external
force  in the right--hand side. Such line has, at least, one proper
curvature determined by the force,  see a physical oriented
exposition of the question by  Synge (1960).
Respectively,  these exterior fields of curvatures of the coordinate
lines enter into   quantum theory.

Thus, the class of the Riemannian coordinates turns out to be a preferred
one. It seems to contradict  the dogma of general relativity on
equivalence of possible systems of coordinates.
The contradiction may possibly  be solved as follows.
A quantum--mechanical description of a physical  system should include
an indication of the way of measurement (observation) of properties
of the system; for a recent discussion of the  question see Rovelli
(1996). In the \Schr, a system of coordinates $\{\xi^i\}$  plays two roles
simultaneously. On the one hand,  it arithmetizes ("digitzes")
the configurational space by its local map  on $R_n$. On the other hand,
it specifies  $n$  primary observables represented  in  quantum mechanics
by  the operators $\hat \xi$ the spectra of which  may be considered
as  formalization of indications of  an apparatus detecting a
 position of the particle.  Numerical values of the indications
should not depend on the arithmetization of \rn and,  in this sense,
should be represented by scalars with respect  to transformations
of $\xi'$s. Therefore,  let us separate the two
roles of  the coordinates as follows: keep for  the
arbitrary coordinates $\xi^i $ the role of arithmetization of  \rn
and introduce  $ 2n $ canonically conjugate scalar functions
$q^{(i)}(\xi), \  p_{(j)}(\xi, p)$  by  the following canonical
transformation:
\beq
\{\xi^i,\ p_j\} \longrightarrow   \{q^{(k)} (\xi), \ p_{(l)} (\xi, p)\}.
\nde
Here   $q^{(k)} (\xi), $ are  fixed $ 2n $ functions such that
$\ {\rm rank}\|\ptl_i q^{(j)}\| = n$,  and
$ p_{(l)} (\xi, p) \defst K_{(l)}^i (\xi) p_i $ where
\beq
 K_{(j)}^i (\xi) = {\rm det}\|\ptl_k q^{(l)}\|    \om^{-\frac 12}
\ep^{ii_2 ... i_n}\,
\ep_{(j j_2 ... j_n)} \ptl_{i_2} q^{(j_2)}.... \ptl_{i_n} q^{(j_n)}
\label{K}
\nde
are  $n$ vector fields and
$\ep^{i_1 i_2 ... i_n},\  \ep_{(j_1 j_2 ... j_n)}$ are completely
antisymmetric symbols  for both  upper and lower indices.
 Of course, one may take     $q^{(i)}(\xi) \equiv  \xi^i$  as  a
particular case, which means that the arithmetization of \rn
and observation of the  particle position  are performed by the same
tools.

The operators in   $L^2 (V_n;\, {\Bbb C};\,  \sqrt \omega\, d^n \xi)$,
corresponding to the scalar primary observables   \\
$q^{(i)}(\xi), \ p_{(j)} (\xi) $  are
\newpage
\beqa
\hat q^{(i)}(\xi) &=&   q^{(i)}(\xi) \cdot {\bf \hat 1}  \label{qq}\\
\hat p_{(i)}  & =&  - \im \hbar\ \left(K^l_{(j)} (\xi) \nabla_l
 + \frac{1}{2} \nabla_l K^l_{(j)}  (\xi) \right).    \label{pk}
\ndea
Introduce a scalar Hamilton operator $\hat H_0^{(\nu)'}(\xi)$ from
the condition that it coincides  with  $\hat H_0^{(\nu)}(\xi)$  when
$q^{(i)}(\xi) \equiv   \xi^i$. Restricting for brevity to the case
of  $ \nu = 2 $, one has
\beqa
  \hat H^{(2)'}_0  &\defst & \frac 1{2m} \hat p_{(i)} \ptl_k q^{(i)}
\om^{kl} \ptl_l  q^{(j)}\hat p_{(j)}\ =
\ - \frac{\h^2}{2m}\left( \Delta_{(\om)}  - \frac 12 \nabla^k v_k +
\frac 14  v^k v_k, \right) \label{ham'}  \\
 v_k &\defst & K_{(i)}^m \nabla_m \ptl_k q^{(i)}.
\ndea
The quantum potential  in  the right--hand side  of (\ref{ham'})
does not depend on the choice of coordinates $\xi^i$,  but  does on
the choice of the observables of  position $q^{i}(\xi)$. This
corresponds to the concept {\it relational quantum mechanics}  developed
by Rovelli (1996) according to which different methods of  observation
of a quantum system give  different amounts
of information on the system.   One may think that choosing the Riemannian
coordinates $y^a$ as observables, i.e.,  $q^{(a)}(\xi)\equiv y^a (\xi) $,
gives maximal information on  the quantum analogue of
the particle moving along a geodesic line in \rn because,  in this
case, no outside information is added in the form of the proper
curvatures of  coordinate lines.\\

{\large\bf 6. ON QFT-APPROACH TO QUANTUM MECHANICS IN CURVED SPACES}\\

{\bf 6.1. QUANTUM FIELD THEORETICAL BASIS }\\

To give a more complete exposition  of the problem of quantum mechanics
in \rn,  an approach     which is  an  alternative
 to  quantization of the geodesic motion will be outlined in the
present section;  details can be found in Tagirov (1999). It
was mentioned  in Section 1  as  the QFT-approach.

The approach is based on quantum theory of the linear real scalar field
$\varphi (x), \   x \in \rinf $ in \rin,
of which the quanta are supposed  to be the point-like spinless and
chargeless particles. Thus, one may think that it should have
a domain of intersection with the  approaches based on quantization
of the geodesic motion and considered in the preceding sections.

A sufficiently general  equation of the field in \rin is
 \begin{eqnarray}
\Box\varphi + \zeta\, R_{(g)}(x)\, \varphi  + \left(\frac{mc}{\h}
\right)^2
 \varphi &=& 0,  \quad x\in \rinf \label{r} \\
\Box \stackrel{def}{=}  g^{\alpha\beta}\nabla_\alpha \nabla_\beta.
\qquad & &\nonumber
\end{eqnarray}
Here $\zeta$ is  an arbitrary dimensionless real constant,
 $ R_{(g)}(x) $  and $\nabla_\alpha$ are, respectively, the scalar curvature  and
the covariant derivative  in  \rin. Two values  of  $\zeta$ are
especially distinguished.  For $\zeta=0$,  the field  $\varphi (x) $
interacts with  the external gravitational field  $g_{\al\be} (x)$
minimally, that is  switched on by immediate  substitution
of the partial derivatives with respect to the Cartesian coordinates
in the standard Klein--Gordon equation by the covariant derivatives
with respect to \rin. For $\zeta = (n-1)/4n $, the interaction is
conformal--invariant in the limit of  $m = 0 $; for $n = 3 $ this property
was first noticed  by R. Penrose (1963)  and  studied in detail
by Chernikov and Tagirov (1968) and Tagirov (1973). The latter authors
had found some other properties of the  theory with the conformal coupling,
which are favorable from the physical point of view.

In  the globally static \rin, by which the scope of the
present article is restricted,  one has
$R_{(g)} = - R_{(\om)}$.  If, in addition,  $ n = 3$, then
$\zeta = 1/6 $ and a nonrelativistic  limit of (\ref{r}) will be the
\Sche just with the Hamiltonian $\hat H^{({\rm DW})}_0 $.
However,  an almost methaphysical question arises here: Why the
quasiclassical approximation leads to the Hamiltonian
 $\hat H^{({\rm DW})}_0$  to  $\zeta = 1/6 $ for any dimension $n$
while the scalar field theory
with the conformal coupling  leads to the same Hamiltonian  only
for the dimension of the real world  $n=3 $?

For a time being, we  consider again the general metric (\ref{g}),
not necessarily  the globally static one.
Canonical quantization  of $\vp (x) $ in the general  \rin   in  the Fock
representation is essentially based  on the  complexification
$\Phi_c\,= \,\Phi \otimes {\Bbb C}$ of  space  $\Phi $ of solutions
to equation (\ref{r})
$\Phi_c\,= \,\Phi \otimes {\Bbb C}$ and
 a selection of    $\Phi_c^\prime \subset \Phi_c $  which   can be
represented as
\begin{equation}
    \Phi_c^\prime\,=\,  \Phi^- \oplus  \Phi^+ , \label{c}
\end{equation}
see, for instance, Gibbons and Pohle (1993).
Here, $  \Phi^- $ and $\Phi^+ $   are  mutually complex conjugate
subspaces of $\Phi_c$,  for  which the
conserved  sesquilinear form
\begin{equation}
\{ \varphi_1, \,\varphi_2 \}_\Sigma \defst \im
 \int_\Sigma d\sigma^\alpha
\left(\ov {\varphi_1}(x)\,\ptl_\alpha \varphi_2 (x)\
- \, \ptl_\alpha \ov {\varphi_1}(x)\,{\varphi_2}(x)\right),
\label{spr}
\end{equation}
is respectively positive and negative, and thus provides
$ \Phi^- $ and $ \Phi^+ $  with  pre-Hilbert structures.

Assume that  a formal (and auxiliary) basis
$\{\varphi (x;\, A)\} $  in $\Phi^- $ exists, which is enumerated by
a multi--index $A$ having values on  a set $\{A\}$
with a measure $\mu (A)$ and  orthonormalized with respect to the inner
product (\ref{spr}). Then, the quantum field operator in a Fock space
${\cal F}$  can be introduced
\begin{equation}
 \check \varphi (x) = \int_{\{A\}} d\mu(A) \left(\check c^+(A)\,
\ov\varphi (x;\,A)\, +\,
 \check c^-(A)\, \varphi (x;\,A)\right)\, \equiv \,\hat\varphi^+ (x)
\,+ \,\hat\varphi^- (x).
  \label{a+}
\end{equation}
(Here and further, operators  in ${\cal F}$  are denoted
 as $\check O$  and   called {\it QFT-operators} contrary
to the quantum--mechanical ones,
or {\it QM-operators}, which are  denoted throughout the paper as
$\hat O$.) The operators
$\check c^+(A)$ and $ \check c^-(A) $ are  creation and annihilation
of the  field modes $\varphi^- (x;\, A) \in \Phi^- $
(or, of {\it  quasi-particles}). They  satisfy the canonical
commutation relations
$$
[\check c^+(A),\,\check c^+(A')] = [\check c^-(A),\,\check c^-(A')] = 0,\
\int_{\{A\}} d\mu(A)\, f(A)\, [\check c^-(A),\, \check c^+(A')] = f(A')
$$
for any appropriate  function $f(A)$. They act  in the Fock space $\cal F$
with the cyclic vector $ |0> $ ({\it the quasi-vacuum}) defined by the
equations
\begin{equation}
 \check c^-(A)\, |0> = 0.  \label{vac1}
\end{equation}

{\bf 6.2. QFT-OPERATORS OF  OBSERVABLES}\\

Now, the following   diffeoinvariant quantum field observables can
be naturally introduced.
{\it The QFT-operator of a number  of quasi-particles}
is determined in the standard way, see, for instance, Schweber (1961),
 Chapter 7, Section 3:
\begin{eqnarray}
\check {\cal  N}(\hat \varphi;\,\Sg) &\defst&
 \int_\Sg d\sigma^\alpha\, (\hat\varphi^+\, \ptl_\alpha \hat\varphi^-
- \ptl_\alpha \hat\varphi^+\, \hat\varphi^-)   \nonumber \\
&\defst& \int_\Sg d\sigma (x) \check N (x), \label{n}
\end{eqnarray}
where $  d\sigma (x) \defst \sqrt{\om(t, \xi)} d^n \xi $
  is  the inner volume element of $\Sg$.

{\it The QFT-operator of the projection of momentum of the field
$\hat\varphi (x)$
on a given vector field $K^\alpha (x)$ } is also a standard expression
determined by the  general--relativistic Lagrangian for $\vp$:
\begin{equation}
\check  {\cal P}_K (\check \varphi; \, \Sg ) \, \defst \,  :\int_\Sg
d\sigma^\alpha \ K^\beta T_{\alpha\beta} (\check \varphi):\ , \label{tk}
\end{equation}
where  the colons denote  the normal product of operators
$\check c^\pm (A) $
and $T_{\alpha\beta} (\varphi) $ is the metric energy--momentum tensor
of the field $\vp (x)$.\\

{\it The $n$ QFT-operators
\begin{eqnarray}
\check {\cal Q}^{(i)}\{\check \varphi;\, \Sg\} & = & i
 \int_\Sg d\sigma^\alpha (x)\ q_\Sg^{(i)}(x)\, \left(\check \varphi^+(x)\
\ptl_\alpha \check \varphi^- (x)\,
- \,\ptl_\alpha \check \varphi^+ (x)\,\check \varphi^- (x)\right) \nonumber \\
&\equiv & \int_\Sg d\sigma (x)\, q_\Sg^{(i)}(x)\, \check N (x).\label{Q2}
\end{eqnarray}
of position of the quasiparticle on $\Sg(t)$ }
observed by means  of  three spatial coordinate  scalar functions
$q_\Sg^{(i)} (x)$  which satisfy the conditions
\begin{equation}
 \ptl^\alpha\Sg \  \ptl_\alpha q_\Sg^{(i)} =  0 , \qquad
\mbox{rank}\|\ptl_\alpha   q_\Sg^{(i)}\| = n,    \label{q}
\end{equation}
and thus define a point on a given  Cauchy hypersurface
$\Sg = \{ x\in\rif\,|\, \Sg(x) = const\}$. In the globally static \rin,
their restrictions on  a completely geodesic  hypersurface
$\Sg$ are just  functions  $q^{(i)}(\xi)$  that have been introduced
in Section 5. (For the Cartesian coordinates in \eu, such an operator  was
considered  by Polubarinov (1973).)  It is  easy to see that
QFT-operators $\check{\cal Q}^{(i)}\{\check \varphi;\, \Sg\}$ are   unique
sesquilinear (in $\check \vp^\pm$) Hermitean   forms  in ${\cal F}$,
which can be constructed from
$\check \vp^\pm, \ \ptl^\al \Sg  \ptl_\al\check\vp^\pm,
$ and  do not contain  derivatives of $q_\Sg^{(i)} (x)$.
	   `
The QFT-observables introduced above are evidently sufficient to describe
quantum dynamics of a single quasi-particle if  there is no processes of
quasi-particle  creation  and annihilation as  in the case of
a globally static \rin, or  if these processes can be  neglected.
Such dynamics is just quantum mechanics of a quasi-particle
the space of states of  which is a subspace of ${\cal F}$ consisting
of  the vectors
\begin{equation}
  |\varphi> \defst \{\varphi,\,  \varphi \}_\Sg^{-1/2}
\int_{\{A\}}d\mu (A)\,\{\varphi (.\,; A),\,\varphi (.)\}_\Sg\ \check c^+(A)\,|0>,
  \label{phi}
\end{equation}
determined by  the field configuration
$$
 \Phi^- \,\ni\,\varphi (x)
= \int_{\{A\}} d\mu (A)\,\{\varphi (.\,;\, A),\,\varphi (.)\}_\Sg
\ \varphi (x;\,A).
$$
Obviously $<\varphi | \varphi> = 1$.
The QM-observables  are determined by the
matrix  elements  of the introduced QFT-operators
between two one--quasiparticle  states
$|\varphi_1>$  and $|\varphi_2>$.\\

{\bf 6.3. ONE-PARTICLE STATES AND OBSERVABLES  }\\

There are infinitely many decompositions  (\ref{c}) and they
generate Fock representations of the  canonical commutation relations
of  the quantum field which are   unitarily unequivalent in general.
The main problem is to distinguish a subspace $\Phi^-$
in the space $\Phi_c$ of solutions of the field  equation (\ref{r})
for which
the introduced  formal quantum mechanics of a quasiparticle

corresponds  to the geodesic motion in \rin   and, therefore, may be called
quantum mechanics  of {\it a particle}. In the general \rin, this problem
can be solved  only  as a quasi-nonrelativistic  asymptotic approximation
to the formal scheme, since the formally exact relativistic quantum
mechanics can be constructed only  in the globally static \rin (see below),
Therefore, we take as $\Phi^-$ a space $\Phi_L^-$ of the following
asymptotic in $c^{-2}$  solutions of eq.(\ref{r})
\begin{equation}
\varphi_L (x) \,= \,\sqrt{\frac{\h}{2mc}}\
exp\left(-\im \frac{mc}{\h}\, S_\Sg (x)\right) \, \hat V_L (x) \psi (x).
\label{az}
\end{equation}
The notation here needs  detailed explanations \\

\noindent
 $S_\Sg (x)$ is  {\it a solution of
the Hamilton--Jacobi equation }
\begin{equation}
\ptl_\alpha S_\Sg \,\ptl^\alpha S_\Sg \,=\,1,  \label{hj}
\end{equation}
which   satisfies  the initial conditions
$\left.S_\Sg (x)\right|_\Sg\, = \, 0 $ and
$\left.\bigl(\tau^\alpha (x) \ptl_\alpha S_\Sg (x)\bigr)\right|_\Sg\,>\, 0$
for any time-like vector field $\tau^\alpha (x)$ directed  into the future.
Any  hypersurface $ S_\Sg(x) =  const $, denoted further simply
as $S$, is a level surface of the normal geodesic flow through $\Sg $.\\

\noindent
$\psi (x) $ is {\it a solution  of \Sche}
\begin{eqnarray}
\quad i\h c(\ptl^\al S \ptl_\al  + \frac12 \Box S )\psi (x) \,
&=& \,\left( \hat H^{({\rm ft})}_L + \Oc\right)\, \psi (x), \label{t1} \\
\hat H^{({\rm ft})}_L &\defst&
  \hat H^{({\rm ft})}_0\,+\,\sum_{l=1}^L \frac{\hat h_n}{(2mc^2)^n},
\label{t2} \\
  \hat H^{({\rm ft})}_0\, &\defst& \,-  \frac{\h^2}{2m} \left(\Delta_S (x) -
\zeta R(x) + \left(\frac{1}{2}(\ptl S \,\ptl\Box S) \
+\, \frac{1}{4} (\Box S)^2 \right)\right), \label{h0}
\end{eqnarray}
and ${\hat h_l}$ are differential operators which are determined by certain
recurrence  relations starting with  $H^{({\rm ft})}_0$ and contain
only derivatives along the  hypersurface $S$ ("spatial derivatives"). \\

 $\hat V_L (x)$ is {\it an asymptotical differential QM-operator} along $S$:
\begin{equation}
\hat V_L (x)\equiv {\bf\hat 1} +
\sum_{l=1}^L \frac{\hat v_l}{(2mc^2)^l} + \Oc
\end{equation}
where   the operators $\hat v_l$  are determined  by  the  asymptotic
relation
\begin{equation}
\{\varphi_1,\,\varphi_2 \}_S =
\left(\psi_1,\,\psi_2\right)_S \,\defst
\,\int_S d\sigma \,\ov\psi_1\, \psi_2 \,
+ \Oc, \quad \vp_1,\vp_2 \in \Phi_L^- . \label{psi}
\end{equation}
It provides   $\Phi_L^-$  with the structure of
$L^2 (S; \Bbb C; d\sigma)$
and  $\psi$  by the standard Born probabilistic interpretation
in each configurational space $S$, i.e. $|\psi(x)|^2$  is  the  probability
density to observe  the field configuration  which  may be
called "a particle" at  the point $x$  belonging to the given hypersurface
$S$. At least, this field configuration satisfies an  intuitive  idea
of what is the quantum particle  as  a localizable object.

Thus,  we have defined the space of states of  a particle  and
can calculate the asymptotical  one--to--one--particle transition
probability amplitudes of form  $<\vp_1| \check {\cal O} |\vp_2>$  for
the QFT-operators  of  the   observables defined above. To this end,
each time when  "the time derivative" $\nabla^\al S \nabla_\al$ appears
  in calculations   it  should be  substituted
by  the   differential operator along $S$ of the apppropiate order
 determined by  the \Sche (\ref{t1}).  In effect,
this completes the deduction  of   quantum mechanics
of the particle in \rin from quantum field theory in the
quasi-nonrelativistic approximation because we have the matrix elements
of observables of the particle, which were considered  in Section 2.
However, to compare the QFT-results with  those
of  the canonical quantization,  we need the operator representations
of the observables as differential operators  in  $L^2 (S; \Bbb C; d\sigma)$.
They are defined up to an  asymptotically unitary transformation
 by the following  general relation:
\begin{eqnarray}
<\varphi_1|  \check {\cal O}|\varphi_2> & = &
\left(\psi_1,\,(\hat O)_L\,\psi_2\right)_S \,\defst
\,\int_S d\sigma \,\ov\psi_1\, (\hat O)_L \psi_2 \,
+ \Oc, \quad \vp_1,\vp_2 \in \Phi_L^-  \label{vv}\\
(\hat O)_L &\defst& (\hat O)_0 +\,\sum_{l=1}^L \frac{\hat o_l}{(2mc^2)^l}
 \label{O1}
\end{eqnarray}
where $\check{\cal O}$ is any  of the QFT-operators and,
again, $\hat o_n$  are  differential QM-operators along $S$ determined
by recurrence relations starting with $(\hat O)_0$.  From eq.(\ref{vv}),
it follows that
\begin{equation}
 (\hat N)_L = {\bf \hat 1} +\Oc.
\end{equation}
For other observables, from eqs.(\ref{Q2}), (\ref{tk}) and (\ref{O1})
one has the following nonrelativistic QM-operators for further
calculations of relativistic corrections:
{\it the particle position on} $S$
\begin{equation}
(\hat q_S^{(i)}(x))_0  =  q_S^{(i)}(x)\, \cdot {\bf \hat 1}, \label{Qi}
\end{equation}
{\it the particle momentum along} $K^\al_{(j)} = \{0,\, K^i_{(j)}\}$ where
$K_{(j)}^i $ is defined as in eq.(\ref{K})
\begin{equation}
(\hat p_{(j)})_0 (x) \defst (P_{K_{(j)}})_0 =
i \hbar\ \left(K^\al_{(j)} \nabla_\al
 + \frac{1}{2} \nabla_\al K^\al_{(j)}  \right),   \label{PKj}
\end{equation}
and  {\it the particle energy}
\begin{equation}
(E (x))_0 \defst  (P_K)_0 \bigr |_{K^\al = c \ptl^\al S}
=   \hat H^{({\rm ft})}_0.
\end{equation}
It is remarkable that not only nonrelativistic expressions for the energy
QM-operator originated by  the energy--momentum tensor $T_{\al\be}$
and  for the
Hamiltonian in the \Sche (\ref{t1}) coincide  but also  their asymptotic
representations of any order $L$  are {\it asymptotically
unitary equivalent}, see Tagirov (1999).\\

{\bf 6.4.  QUANTUM  MECHANICS IN THE  GLOBALLY STATIC SPACE-TIME  AND
DEFORMATION OF CANONICAL COMMUTATION RELATIONS}\\

Pass now to the case  of globally static \rin  and  consider  it, as in
Sections 2 -- 5 ,  in  a
system of coordinates $\{x^\al\} \sim \{t, \xi \}  $ in  which
$ \om_{ij}(t, \xi)  \equiv  \om_{ij}(\xi). $ Then,  the  asymptotic
expansions above  can be  represented  in the  formal closed forms
\begin{eqnarray}
\hat H^{({\rm ft})}_\infty &=&  mc^2 \left(\left(1 +
\frac{2\hat H^{({\rm ft})}_0}{mc^2}\right)^{1/2} - 1 \right),
\qquad  \hat H^{({\rm ft})}_0 = -\frac{\h^2}{2m}(\Delta_S - \zeta \,R),
\label{hinf}\\
 \hat V_{\infty} &=&
\left(1 + \frac{2 \hat H^{({\rm ft})}_0}{mc^2}\right)^{-1/4}, \label{vinf} \\
({\hat p}_{(j)})_\infty (x)  &=& \,-\frac{i\h}{2}\,
 \hat V_\infty^{-1}\cdot (K^i_{(j)}\nabla_i)\cdot \hat V_\infty \, + \,
\frac{i\h}{2}\, \hat,V_\infty \cdot (K^i_{(j)}\nabla_i)^{\dagger}
\cdot\hat V_\infty^{-1}
\nonumber\\
 & - &  \frac{\h\zeta}{2mc^2}\,\hat V_\infty
\cdot (c\ptl^\al S \nabla_\al) (\nabla_r K^r{(j)})\cdot
\hat V_\infty \label{pinf}\\
c\ (\hat p_{\ptl S})_\infty (x)
&=& mc^2 \left(1 + \frac{2\hat H^{({\rm ft})}_0}{mc^2}\right)^{1/2},
\qquad ({\rm energy\ operator})\label{einf}\\
(\hat q_S^{(i)})_\infty  &=& q_S^{(i)} (x)\, + \, \frac{1}{2}
\left[ [\hat V_{\infty},\ q_S^{(i)} (x)],\ \hat V_\infty^{-1}\right].
  \label{qinf}
\end{eqnarray}
Recall that we use $\nabla_\al$ and $\nabla_i$ to denote
the covariant derivative with respect to the metric tensors
$g_{\al\be}$ and $\om_{ij}$ respectively.\\

{\bf 6.5.   QFT-APPROACH VS. QUANTIZATION OF GEODESIC MOTION }\\

What conclusions can be made  from the formulae obtained  just now
comparing them with the results of Sections 2 and 3? \\

1. They are diffeoinvariant in \rn and \rin owing to  introduction
of the functions $q_S^{(i)} (x)$ which were proposed  in Section 5  to
separate  background coordinates $\xi$ on \rn (that is on $S$)  from
the coordinates $q_S^{(i)}$ in terms of which  a position  on $S$ of the
quantum particle is observed.

2. The relativistic  Hamiltonian $\hat H^{({\rm ft})}_\infty$ is expressed
through the  non-relativistic one $\hat H^{({\rm ft})}_0$  just  by
formula (\ref{hamn}) and  supports the asymptotic meaning of
quantization of $H (p, \xi)$, eq.(\ref{ham}).

3. The nonrelativistic Hamiltonian  $\hat H^{({\rm ft})}_0$  is  similar
to  DeWitt's one $\hat H^{({\rm DW})}_0 (\xi)$, eq.(\ref{hamR}), but
the coefficient before the scalar curvature $R$  is an arbitrary
constant $\zeta$ in $\hat H_0^{({\rm ft})}$ instead of value $(1/6)$  in
$\hat H_0 (\xi)$. As it has already been said, the latter
distinguished  value of $\zeta$  corresponds to the conformal coupling of
$\vp$ to gravitation, but only when  $n =3$.  Another interesting
difference is that  $\hat H^{({\rm ft})}_0$ is an exact expression
 with  no other quantum potential terms if $c^{-1}  = 0$ while
$\hat H^{({\rm DW})}_0 (\xi)$ is  the  quasi-classical  approximate
expression. This difference  is very interesting and  poses
a question:  the non-diffeoinvariant part of quantum potential, is
it a deficiency of the  quantization of mechanics  or is  its absence
in the QFT-approach   a manifestation of some
incompleteness of the canonical quantization of  field?

4. The most remarkable consequence of the QFT-approach
is that the position operators $(\hat q_S^{(i)})_L$ do not commute among
themselves except the case of $L=0$ and the same takes place for
the momentum ones $({\hat p}_{(j)})_N$.  Therefore, the canonical
commutation  relations  (\ref{ccr})  are fulfilled only in the exactly
non-relativistic case $c^{-1}  = 0$ and  {\it the quasi-nonrelativistic
commutation relations of primary observables  are a deformation of the
nonrelativistic ones}. An analogous deformation of the  ${\bf o(3)}$
algebra of the spin $1/2$ operators  arises  when the
QFT-approach is used for the Dirac particles (Tagirov, 1996).

5. The QFT-approach   gives  at once
a quasi-nonrelativistic quantum mechanics in the general \rin and
arbitrary  frame reference formed  by  the normal  geodesic flow
through an  arbitrary Cauchy hypersurface $S$. In contrast,
 quantization of mechanics is formulated
above   only  in the  globally static  and topologically elementary
\rn  and  only in the frame of reference in which  the metric tensor  is
time--independent; this frame  is formed by the Killing flow.
A consecutive  formulation in the the latter case needs a special study.

6. Morever,  since  the theory is formulated actually  in  terms of
matrix elements of the form  (\ref{O1}), it may be applied  to   \rn  of any
topology.

7. The QFT--approach  opens a way to  formulate
quantum mechanics'  of particles  with non-zero spin, for which there are
no  classical counterparts. \\

{\bf 6.6.  SPACE QUANTIZATION IN THE FRIEDMANN--ROBERTSON--WALKER
UNIVERSE}\\

At last,  I should like to announce  an interesting result   obtained
in the QFT-approach  for  the case of  the
Friedmann--Robertson--Walker universe  and  a natural frame of
reference in it, in which
\beq
ds^2 =  c^2 dt^2 - b^2 (t) \tilde\omega _{ij}(\xi)  d\xi^i d\xi^j,
\quad i, j,...= 1, 2, 3.
\nde
Let
$ q_{S\sim t_0}^{(i)} (\xi,) = X^{(i)} (\xi)$  be the
normal Riemannian coordinates which are measured in the  units of
the cosmological scale factor $b(t)$ .  In   the standard Euclidean
 vector notation\\
$ \{X^{(1)}, X^{(2)}, X^{(3)}\}\equiv \stc{\to}{X}$, see, for instance,
Weinberg (1972), Chapter 13, one has
$$
\tilde\omega _{ij} d\xi^i d\xi^j\,  = \,
\left(d \stc{\to}{X}\right)^2 +
\frac{k\left(\stc{\to}{X}\cdot d\stc{\to}{X} \right)}
{1 - k {\stc{\to}{X}}^2}
$$
where $ k= 1, \quad 0, \quad -1 $
for the spatially spherical, flat and hyperspherical universes, respectively.

Since the  space geometry depends on  the cosmological time $t$, the
structure  of quasi-non-relativistic quantum mechanics and, thus,
the notion  of a  particle  are specified by   an initial
moment $t_0$ at which the Cauchy problem  for the \Sche  is  posed.
It is remarkable that, for  $O(c^{-2})$   any coordinates
$q^{(i)} (\xi)$, the first nonvanishing relativistic correction  to
$(\hat q_{S\sim t_0}^{(i)})_L (\xi)$  is  of order  $O(c^{-4})$. For  the
normal Riemannian  coordinates, one has
\beq
\left[(\hat X^{(i)})_2, \ (\hat Y^{(j)})_2 \right]
= - k \left(\frac{\lambda_C}{b(t_0)}\right)^4
\left(X^{(i)}\frac{\ptl}{\ptl Y^{(j)}} -
 Y^{(i)}\frac{\ptl}{\ptl X^{(j)}}\right)
+ O(c^{(-6)})  \label{sny}
\nde
where $\lambda_C = \h/mc$ is the Compton wave length of the particle. .
It is  remarkable that (\ref{sny}) is the  ${\bf o(3)}$--part
of the basic  formula in Snyder's theory  of  quantized
Minkowsky space-time Snyder (1947):
$$
\left[(\hat X^\al)_2, \ (\hat Y^\be)_2 \right] = l_0^2 L^{(\al\be)}.
$$
where  $L^{(\al\be)}$ are the Lorentz group generators and $l_0$ is an
elementary length.    According to (\ref{sny}), {\it the space
seems to be quantized  in principle  in  the  standard theory} with
no additional hypotheses,
except  the case of spatially flat universe ($k=0$). The elementary
length   $l(t_0)= k (\lambda_C/b (t_0)) \lambda_C$
depends on the moment  of time in which the \Schr is specified. However,
one  should remember  that  a particle specified by the moment $t_0$
has to  be  sufficiently  heavy  the quasi-nonrelativistic  approximation
to be  valid  and  the processes  of   particle creation and annihilation
caused by the time --dependence of the cosmological factor $b(t)$
to be negligible. To conclude finally,   may  the  space quantization
have  at least, a hypothetical physical sense,  or, is it  an artefact
of the approximation, it is necessary also to  estimate  contributions
of the next order in $c^{-2}$  and, in the case of $k = 1$,
to take into account singularity at ${\stc{\to}{X}}^2 = 1$. The present
author hopes  to  present such study in future elsewhere.\\

The  most  important  point is,  however,   that  the deformation of the
canonical commutation relations and, consequently, the space quantization
disappear for any  $L$ and $t_0$ in the  exceptional case $k = 0$, i.e.
in  the spatially
flat universe, see the general proof  in Tagirov (2000) .  This  fact
correlates remarkably with that, according to  the modern astrophysical
observational data, the  real Universe is  spatially flat with
fantastically high  accuracy which needs to be explained (the so called
{problem of  flatness}). Couldn't the flatness  have a relation to
that the spatially-flat FRW  models are discretely distinguished by
the first principles of  quantum--mechanics? \\
\newpage
\noindent
{\bf ACKNOWLEDGEMENT}\\

The present study is supported in a part by the Russian Foundation for
Fundamental Research, Grant No 00-01-00871.\\

The author is grateful  to Dr. A.A.Sharapov (Tomsk  State University) and
Dr. A.A.Vladimirov (Joint Institute for Nuclear Research, Dubna) for
useful references  and explanations. \\

\noindent
{\bf REFERENCES}\\

\noindent
 Abraham, R. and   Marsden,  J.E. (1978) {\it  Foundations of Mechanics},
Reading: Ma. Benjamin/Cummings Publications Co.\\
Agarwal, G.S. and  Wolf, E. (1970). Phys.Rev.  {\bf D2},  2161.\\
Bayen, F. {\it et al}.(1978), {\it  Annals of Physics (N.Y.)},
{\bf 111}, 111.\\
 Berezin, F.A. and Shubin, M.A. (1984). {\it Schr\"odinger Equation}.(In
Russian),  Moscow: Moscow University Press.\\
Bordemann, M.,  Neumaier, and  Waldmann S. (1998).
Communications on Mathematical Physics, {\bf 198},  363.\\
Chernikov, N.A. and  Tagirov, E.A. (1969)  Annales de l'Institute
Henry Poincar\`e, {\bf A9}, 109.\\
Gibbons,  G.W. and  Pohle, H.J.  (1993).{\it Nuclear Physics},
{\bf B410},  117.\\
Gavrilov, S.P. and  Gitman, D.M. (2001).  {\it  International Journal
of Modern Physics}, {\bf A15},  4499. \\
Gitman, D.M. and Tyutin, I.V. (1990).
{\it Classical and Quantum Gravity},  {\bf 70}, 2131.\\
 Destri, C.,  Maraner, P. and   Onofri, E. (1994). {\it Nuovo Cimento},
{\bf A107},  237. \\

DeWitt, B.S. (1957). {\it Reviews of Modern Physics}, {\bf 29}, 377. \\
 Dirac, P.A.M. (1948). {\it Principles of Quantum Mechanics},
Cambridge University Press, Cambridge.\\
 D'Olivo, J.C.  and  Torres, M. (1989).
{\it Journal of Physics A: Math. Gen.},  {\bf 21}, 3355. \\
Fedosov, B.V. (1994). {\it Journal of Differential Geometry},
{\bf 40},  213.\\
Groshe, C., Pogosyan, G.S. and Sissakian, A.N. (1997).
{\it Particles and Nuclei (Dubna)}, {\bf 28}, 1229. \\
Kalinin, D. (1999). {\it Reports on Mathematical Physics}, {\bf 43}, 147. \\
Karasev, M.V. and  Maslov,  V.P. (1991). {\it   Nonlinear
Poisson Brackets. Geometry  and Quantization} (in Russian), Nauka, Moscow.
(1993). (in English) American Mathematical Society, RI.\\
Marinov, M.S. (1995). {\it Journal of Mathematical Physics}, {\bf 36},  2458.
 McLaughlin, D.W. and   Schulman, L.S. (1971). {\it Journal of Mathematical
Physics}, {\bf 12}, 2520.\\
Moyal. J.E. (1949). Proceedings of the Cambridge Philosophical
Society, {\bf 45}, 99.\\
Ogawa, N., Fuji, K. and   Kobushkin, A. (1990)  Progress of
Theoretical Physics, {\bf 83}, 894.\\
Pauli, W. (1950--1951) {\it Feldquantisierung. Lecture notes.} Zurich. \\
Penrose, R. (1964). In {\it Relativity, Groups and Topology}, B.DeWitt, ed.,
Gordon and Breach, London. \\
Rivier, D.C. (1957). {\it Physical Review},  {\bf 83}, 862.\\
Polubarinov, I.V.  (1973). {\it  JINR Communication P2--8371}, Dubna. \\
Rovelli, C. (1996). {\it International Journal of Theoretical Physics},
{\bf 35}, 1637. \\
Rovelli, C. (1999). Preprint  hep-th/9910131.\\
Schweber, S. (1961). {\it An Introduction to Relativistic Quantum Field
Theory}, Row, Peterson and Co., N.Y. \\
 \'Sniatycki, J. (1980). {\it Geometric Quantization and Quantum
Mechanics},  Springer--Verlag, New York, Heidelberg,  Berlin. \\
Snyder, H. (1947) {\it Physical Review}, {\bf 71}, 38. \\
Sudbery, A. (1986). {\it  Quantum Mechanics and Particles
of Nature}, Cambridge: Cambridge University Press. \\
 Synge, J.L. (1960). {\it  Relativity: The General Theory},
North--Holland Co., Amsterdam. \\
Tagirov, E.A. (1973). {\it Annals of Physics (N.Y.)}, {\bf 76}, 561. \\
Tagirov, E.A. (1990). {\it  Theoretical and Mathematical Physics}, {\bf  84},
No 3, 419.\\
Tagirov, E.A. (1992). {\it  ibid}, {\bf  90}, No 3, 412.\\
Tagirov, E.A. (1996). {\it  ibid}, {\bf  106}, No 1, 122.\\
Tagirov, E.A. (1999). {\it Classical and Quantum Gravity}, {\bf 16}, 2165.\\
Tagirov, E.A. (2000). In {\it  Proceedings of the International Workshop
"Hot Points in Astrophysics"} held in JINR, Dubna, 22 -- 26 August 2000,
383;     preprint    gr-qc 0011011. \\
Von Neumann, J. (1955). {\it Mathematical Foundations of Quantum
Mechanics}, Princeton University Press, Princeton.
Van Vleck, J.H. (1928). {\it  Proceedings of the National Academy
of Sciences of USA}. {\bf 14}, 178.\\
Veblen, O. (1927).  {Invariants of Quadratic Differential
Forms}, Cambridge: Cambridge University Press.\\
Weinberg, S. (1972) {\it  Gravitation and Cosmology}, John Wiley and
Sons, N.Y.\\
Weyl, H. (1931). {\it  The theory of groups  and quantum mechanics},
Dover Publications, N.Y.\\

\newpage
{\bf Footnotes}\\

\footnotemark[1]The term "deformation" is used very deliberately
in the present  paper to denote a substitution of the Poisson or Lie
brackets  by an asymptotic sum the terms of which are bilinear and
antisymmetric  in the same sense as the brackets themselves are;
this is only one of the properties of the notion  of deformation
used in the mathematically more rigorous texts.\\

\footnotemark[2]Here and  further the spaces  $L^2 (\rif)$ and
$L^2 (\rnf)$
are  defined  with respect to the natural measures
induced by the  corresponding Riemannian metric forms. This  allows
to consider the functions from these spaces as scalars with respect to
the diffeomorphisms in \rin and  \rn. If  there were  no
metric, a more complicate  construction  with  a class of the equivalent
Lebesgue measures on the configurational space and the half--forms
 instead of the scalars  should  be used, see
Abraham and Marsden (1978), p.427.\\

\footnotemark[3]It is  well-known
that the operators
$\hat\xi^i,\ \hat p_j$ are  symmetrical, or, {\it Hermitean},
but not self-adjoint ones in $L^2 (\rnf;\ {\Bbb C};\ \sqrt \omega\ d^n \xi)$.

A consecutive solution of this problem is achieved  by introducing of
the rigged Hilbert space, see, for instance, Sudbery (1986). Here, we shall adopt a
more simple assumption that only an appropriate dense subset in
$ L^2 (\rnf; {\Bbb C};  \sqrt \omega\ d^n \xi) $ is under consideration.\\

\end{document}